\documentstyle[12pt,graphicx]{article}
\title{Yang-Mills condensate dark energy
                 coupled with matter and radiation}
\author{\small    Y. Zhang\cite{email}, T.Y.  Xia ,
                            and W. Zhao \\
        \small Astrophysics Center \\
       \small University of Science and Technology of China \\
       \small Hefei, Anhui, China }
 \date{}

\begin{document}
\maketitle
\baselineskip=19truept
\def\vek{\vec{k}}

\newcommand{\be}{\begin{equation}}
\newcommand{\ee}{\end{equation}}
\newcommand{\ba}{\begin{eqnarray}}
\newcommand{\ea}{\end{eqnarray}}

 \sf

\begin{center}
\Large  Abstract
\end{center}
\begin{quote}
{
The coincidence problem is studied for the dark energy model of
effective Yang-Mills condensate (YMC)  in a flat expanding universe
during the matter-dominated stage.
The YMC energy  $\rho_y(t)$ is taken to represent the dark energy,
which is coupled either with the matter $\rho_m(t)$, or
with both the matter and the radiation components $\rho_r(t)$.
The effective YM Lagrangian is completely determined
by quantum field theory
up to 1-loop order with an energy scale
$\kappa^{1/2}\sim 10^{-3}$ev as a model parameter,
and for each coupling, there is an extra model parameter introduced.
Beyond the non-coupling case,
we have extensively studied four types of coupling models:
1)  the YMC decaying into the matter
at a rate $\Gamma \sim 0.5H_0$,
where $H_0$ is the Hubble constant;
2)  the matter decaying into the YMC
at a rate $\Gamma \sim 0.02H_0$;
3)  the YMC decaying into both the matter
(at $\Gamma \sim 0.5H_0$) and the radiation
(at  $\Gamma' \sim 1.8\times 10^{-4} H_0$).
4) both the matter (at  $\Gamma \sim 0.5H_0$)
and the radiation (at  $\Gamma' \sim 1.8\times 10^{-4} H_0$)
decaying into the YMC.
In each of these four models,
we have also explored various couplings.
For all these models and for a variety of functional forms
of the couplings $\Gamma$ and $\Gamma'$,
it is found that the overall feature of the
cosmic evolution for the YMC component is similar.
Starting from the equality of radiation-matter
$\rho_{mi}= \rho_{ri}$,
for generic initial conditions of $\rho_{yi}$ subdominant
by a factor ranging over $8$ orders of magnitude
from $10^{-10}$ to  $10^{-2}$,
the models always have a scaling solution during the early stages,
and the YMC always levels off  at late time
and becomes dominant,
so that the universe transits from
the matter-dominated into the dark energy-dominated stage
at $z\simeq (0.3\sim 0.5)$,
and evolves to the present state with
$\Omega_{y}\simeq 0.7$, $\Omega_m\simeq 0.3$,
and $\Omega_r\simeq 10^{-5}$.
For the matter and radiation components,
their evolutions depend on how they couple to the YMC.
It is found that,
if the YMC decays into a component
with a rate being a constant or depending on $\rho_y$,
as in Model 1 and Model 3,
then in the early stages one has $\rho_m(t) \propto a(t)^{-3}$
and $\rho_r(t) \propto a(t)^{-4}$
like the non-coupling case, but later around $z\sim 0$,
this component also stops decreasing and levels off,
asymptotically approaching to a constant value,
just as $\rho_y(t)$ does.
In these cases, the equation of state (EoS) of
the YMC $w_y=\rho_y/p_y$ crosses over
$-1$ around $z\sim 2$   and
takes the current value $w_y\simeq -1.1$ at $z=0$,
consistent with the recent preliminary observations on
supernovae Ia.
But if the coupling is such that
the matter decays into the YMC in Model 2,
or if both the matter and radiation decay into the YMC in Model 4,
then $\rho_m(t) \propto a(t)^{-3}$  and
$\rho_r(t) \propto a(t)^{-4}$ approximately for all the time,
and $w_y$ approaches to $-1$ but does not cross over $-1$.
We have also demonstrated explicitly that,
the coupled dynamics for ($\rho_y(t)$, $\rho_m(t)$) in  Model 1 ,
or for ($\rho_y(t)$, $\rho_m(t)$, $\rho_r(t)$) in Model 3,
as $t\rightarrow \infty$,
is a stable attractor; in Model 2 and Model 4
the dynamics of $\rho_y(t)$
has a stable attractor as $t\rightarrow \infty$.
Therefore, under very generic circumstances,
the existence of
the scaling solution during the early stages and the
subsequential exit from the scaling regime around
$z\simeq (0.3\sim 0.5)$
are inevitable.
Thus the coincidence problem is naturally solved
in the effective YMC dark energy models.
 }
\end{quote}


email: yzh@ustc.edu.cn

\small
\newpage
\baselineskip=19truept

\begin{center}
{\em\Large 1. Introduction}
\end{center}

The observations on the cosmic microwave background
radiation (CMB)\cite{map} suggests a flat universe.
This observation together with
that from the Type Ia Supernova (SN Ia) \cite{sn}
implies that the universe consists of some mysteries
dark energy ($\Omega_{\Lambda} \sim 0.73$),
dark matter ($\Omega_{d} \sim 0.23$),  and
ordinary baryon matter ($\Omega_{b} \sim 0.04$),
and a radiation component $\Omega_r\sim 10^{-5}$.
This is also supported by
and the large scale structure of the universe \cite{sdss}.
The dark energy as the dominant cosmic energy component
drives the current accelerating expansion of the universe.
The simplest model for the dark energy
is the cosmological constant $\Lambda$,
which corresponds to a homogeneous and time-independent energy density
$\rho_{\Lambda}\sim 5.8 h_0^2\times 10^{-11}$ev$^4$ with
an EoS of
$w_{\Lambda}=p_{\Lambda}/\rho_{\Lambda}= -1$.
Although this model can account for the observation so far,
it has a coincidence problem.
The observations show that
the present value of the energy density for
the matter component $\rho_m=\rho_d+\rho_b$ is
about one third of $\rho_{\Lambda}$,
but it  varies with time as $\rho_m(t) \propto a(t)^{-3}$.
So,  for example,  at an earlier time of radiation-matter equality
with  redshift $z \simeq 3454$ \cite{spergel},
$\rho_{\Lambda}$ should be a very fine
tuned value $\simeq 6.3\times 10^{-11} \rho_m(t) $.
Otherwise,
a slightly variant initial value of $\rho_{\Lambda}$ would lead to
a value of the ratio  $\rho_{\Lambda}/\rho_m$
drastically different from the observed one.
This  is called the coincidence problem.

One class of models aiming at solving the coincidence problem
is based upon the dynamics of
some scalar field $\phi$, such as quintessence \cite{Ratra},
K-essence \cite{Chiba},  tachyon \cite{padmanabhan},
phantom \cite{Caldwell},
and quintom \cite{hu},  etc.
These models can give rise to certain desired features of
evolutional dynamics, such as scaling solutions \cite{Copeland98}
and tracking behavior.
As a common point,
these scalar models need to make use of
some special forms of the potential $V(\phi)$
with certain chosen parameters.
For example, in the  quintessence model,
one needs to choose $V(\phi)\propto (M/\phi)^{\alpha }$
with $\alpha$ and $M$ being a positive number,
or $\propto e^{-\phi/m_{pl}}$ \cite{Zlatev}.
In phantom model one may take
$V(\phi)=V_0[\cosh(\alpha\phi/m_{pl})]^{-1}$ \cite{Singh03}.
Moreover, phantom models typically
introduce a negative kinetic energy term
$-\dot \phi^2 /2$.
Some of these models are expected to be low energy
effective  field theory,
coming from some fundamental field theory,
others are simply introduced by hand.

As far as cosmological  observations are concerned,
a cosmological model has  to give the current status:
$\Omega_{\Lambda} \sim 0.7$ and $\Omega_m \sim 0.3$.
Besides, it is safer for a model not to
contradict the conventional scenario
in the Big Bang model from
the energy scale $\sim1 $Mev down to the present.
Therefore, one idea to solve the coincidence problem
is that during the early stages of the expansion
the dark energy density need not to be a constant,
but varies with time.
Even its EoS $w$ need not to be close
to $ -1$ in early stages.
Only at some rather recent moment has it become dominant and acquired
an EoS $w\sim -1$.
In order to allow the conventional cosmological processes,
such as the nucleosynthesis  and the recombination,  etc.,
to have been occurred in the past,
the dark energy component should be subdominant to
the matter component
early stages.

In the  approach of the scalar field models,
certain coupling
has been introduced between the scalar field dark energy
and the matter \cite{Ellis89} \cite{Amendola00}.
With some particular choice of the  model parameters,
there can exist a scaling solution of dynamics,
in which the dark energy density is proportional to that of matter
during the early stages of evolution.
However, to achieve the tracking solution,
some scalar models  need to have a very large coupling,
so that the universe would enter the acceleration stage
soon after the matter era.
This would result in a picture of structure formation,
totally different from that required by the observations.
To remedy this defect some models \cite{Amendola01} \cite{Valentini}
introduce certain particular form of couplings,
but still the entrance to the accelerating stage
is a little earlier than what observations suggested.
On the other hand, it has been pointed out that the k-essence models
always have, at some stage,  the difficulty of superluminal propagation,
leading to violation of causality,
and the EoS and the sound speed of k-essence
could be greater than $>1$ \cite{Bonvin}.
Besides the unconventional negative kinetic energy,
after the EoS  $w$ cross $-1$,
a class of scalar models may suffer from
severe quantum instabilities and
the Big Rip singularity,
i.e. either the energy density $\rho$, or the pressure $p$,
may grow to infinity within a finite time.
More recently, an overall estimate of scalar models
has been given
on the issue of coincidence problem.
By examining the most general form of scalar Lagrangian
with a generic coupling between scalar
dark energy and dark matter,
it has been shown that
a vast class of scalar models,
including all the models in current literature,
have the difficulty to
implement both a scaling solution without singularity
and a sequence of expansion epochs that required
by standard cosmology,
such as the radiation-dominated, matter-dominated,
and dark energy-dominated epochs \cite{tsujikawa}.
Therefore, the coincidence problem still remains
after over a decade of extensive studies \cite{copeland2006}.

The introduction of the quantum effective YMC
into cosmology \cite{Zhang07}  has been motivated  by
the fact that the $SU(3)$ YMC has given
a phenomenological description
of the vacuum within hadrons confining quarks,
and yet at the same time all the important properties
of a proper quantum field are kept,
such as the Lorentz invariance,
the gauge symmetry, and the correct
trace anomaly \cite{Adler}.
Quarks inside a hadron  would experience the existence
of the Bag constant,  $B$,
which is equivalent to an energy density $\rho= B$
and  a pressure $p= -B$.
So quarks would feel an energy-momentum  tensor
of the vacuum as $T_{\mu \nu}= B \, diag (1,-1,-1,-1)$.
This non-trivial vacuum has been formed mainly
by the contributions from the quantum effective YMC,
and from the possible interactions with quarks.
Our thinking has been that, like the vacuum of QCD inside a hadron,
what if the vacuum of the universe as a whole is also
filled with some kind of YMC.
Gauge fields play a very important role in,
and are the indispensable cornerstone to,  particle physics.
All known fundamental interactions between particles
are mediated through gauge bosons.
Generally speaking, as a gauge field,
the YMC under consideration may have
interactions with other species of particles in the universe.
In our previous studies of  \cite{Zhang, Zhao},
the possible interaction of the YMC with other
cosmic components have not been examined.
However, unlike those well known interactions in QED, QCD,
and the electro-weak unification,
here   at the moment we do not yet have a model
for the details of the microscopic interactions between
the YMC and other particle.
Therefore, in this paper on the dark energy model,
 we will adopt a simple description of the possible interactions
between the YMC and other cosmic particles.
That is, we introduce coupling terms in
the continuity equations of cosmic energy densities,
such as the YMC, the  matter and the radiation,
study  the cosmic evolution of the universe
from the matter dominated era up to the  present.
As shall be seen,
in our model the current status of the universe
turns out to be a natural  result
of evolutional dynamics driven by
the effective YMC as the dark energy,
plus the matter and the radiation that are coupled to the former.
What is important is that
this has been achieved with
a choice of the initial value of the fractional
energy density of the YMC
ranging from $10^{-10}$ to $10^{-2}$.
As a novel feature in contrast to the non-coupling
models,
the coupling YMC dark energy models  can
give rise to an EoS of dark energy $w_y$ crossing $-1$,
say $w_y \sim -1.1$ at present.
The coincidence problem can be solved.

In section 2, as a basis for the setup,
an introduction is given to
the effective Yang-Mills condensate theory,
and the dynamic equations for the three components
in the Robertson-Walker spacetime
are derived.
Section 3 is about the simplest case of non-coupling.
Section 4 studies the dynamic cosmic evolution
with the effective YMC
decaying into the  matter component.
There are two dynamic equations
for the YMC and the matter, respectively.
Section 5 studies
the model in which the matter decays into the YMC.
The matter component has different dynamic evolutions
especially at late stages for these two coupling models.
Nevertheless, in both these two  models,
the YMC has a scaling solution and a tracking behavior
as a natural outcome of the dynamic evolution.
Section 6 extends to the general case
with the YMC coupling to both the matter and the radiation.
Now one has one more coupled dynamic equation
and an extra coupling for the radiation component.
Two cases are studied,
the YMC decaying into the matter and radiation,
and the matter and radiation decaying into the YMC.
The dynamics is examined parallel to sections 4 and 5.
In each of these four models,
various functional forms of coupling have been explored.
The major part of the study is in sections 4, 5,  and 6,
which contain the calculations and the results.
Section 7 gives an analysis on asymptotic behavior
of the dynamic system at $t \rightarrow \infty$.
It is found that there exists a unique attractor
 in the asymptotic
region, which is stable against perturbations.
Section 8 contains a summary and discussions.

Throughout this paper we will work with unit,
 in which
$c=\hbar =k_B=1$.

\begin{center}
{\em\Large 2. YM condensate as dark energy}
\end{center}

In the effective YMC dark energy
model, the effective YM field Lagrangian is
given by \cite{Zhang07} \cite{Adler}:
\be \label{lagrangian}
L_{eff}=\frac{1}{2}bF(\ln|\frac{F}{\kappa^2}|-1)
\ee
where $\kappa$ is the
renormalization scale of dimension of squared mass,
$F\equiv-\frac{1}{2}F^a_{\mu\nu}F^{a\mu\nu}=E^2-B^2$
plays the role
of the order parameter of the YMC.
In this paper, for simplicity, we only discuss the
pure `electric' case, $F=E^2$.
The Callan-Symanzik coefficient
$b=(11N-2N_f)/24\pi^2$ for $SU(N)$
with $N_f$ being the number of
quark flavors.
For the gauge group $SU(2)$ considered in this paper,
one has $b=2\cdot11/24\pi^2$ when the fermion's
contribution is neglected,
and $b=2\cdot 5/24\pi^2$
when the number of quark flavors is taken to be $N_f=6$.
For the case of $SU(3)$ the effective
Lagrangian in Eq.(\ref{lagrangian})
leads to a phenomenological
description of the asymptotic freedom for
the quarks inside hadrons \cite{Adler}.
It should be noticed that
the $SU(2)$ YM field is introduced here as a model for
the cosmic dark energy,
it may not be directly identified as
the QCD gluon fields, nor
the weak-electromagnetic unification
gauge fields, such as $Z^0$ and $W^{\pm}$.
As will be seen later,
the YMC has an energy scale
characterized by the parameter $\kappa^{1/2}\sim 10^{-3}$ ev,
much smaller than that of QCD and
of the weak-electromagnetic unification.
An explanation can be given for
the form in Eq.(\ref{lagrangian})
as an effective Lagrangian
up to 1-loop quantum correction \cite{Adler}.
A classical $SU(N)$ YM field Lagrangian is
\[
L= \frac{1}{2g_0^2}F,
\]
where $g_0$ is the bare coupling constant.
As is known, when the 1-loop  quantum corrections are included,
the bare coupling constant $g_0$ will be replaced
by the running coupling $g$
as the following \cite{Adler}\cite{weinberg}
\[
g_0^2 \rightarrow g^2=
\frac{4\cdot 12\pi^2}{11N\ln(\frac{k^2}{k_0^2})}
=  \frac{2}{b \ln(\frac{k^2}{k_0^2})},
\]
where $k$ is the momentum transfer and $k_0$ is
the energy scale.
To build up an effective theory  \cite{Adler},
one may just replace the momentum transfer $k^2$
by the field strength $F$ in the following manner:
\[
\ln(\frac{k^2}{k_0^2})\rightarrow
2\ln |\frac{F}{\kappa^2 \, e}|
=  2(\ln  |\frac{F}{\kappa^2 } | -1  ),
\]
yielding Eq.(\ref{lagrangian}).
We like to point out that the renormalization scale $\kappa$ is
the only parameter of this effective YM model,
and its value should be determined by comparing the observations.
In contrast to the scalar-field dark energy models,
the YMC Lagrangian is completely
fixed by quantum corrections
up to order of 1-loops,
and there is no room for adjusting its functional form.
This is an attractive feature of
the effective YMC dark energy model.

From Eq.(\ref{lagrangian}) we can derive
the energy density and the pressure of the
condensate \cite{Zhang07, Zhang} in the flat R-W spacetime:
\be \label{rhoy}
\rho_y=\frac{1}{2}\epsilon E^2+\frac{1}{2}bE^2,
\ee
\be \label{py}
p_y=\frac{1}{6}\epsilon E^2-\frac{1}{2}bE^2,
\ee
where $\epsilon$ is called the dielectric constant
of the YMC, given by  \cite{Adler}
\be
\epsilon=2\frac{\partial L_{eff}}{\partial F}
=b\ln|\frac{F}{\kappa^2}|.
\ee
The EoS of YMC is given by
 \be \label{wy}
w_y  =\frac{p_y}{\rho_y}= \frac{y-3}{3y+3},
 \ee
where
\be \label{defy}
y\equiv \frac{\epsilon}{b}=\ln|\frac{E^2}{\kappa^2}|
\ee
is a dimensionless quantity,
in terms of which the energy density and pressure of
the YMC  will be given by
\be \label{rhoyy}
\rho_y= \frac{1}{2}b \kappa^2(y+1)e^y,
\ee
\be
p_y= \frac{1}{2}b \kappa^2(\frac{1}{3}y-1)e^y.
\ee
One sees that,
to ensure that the energy density be positive
in any physically viable model,
the allowance for the quantity $y$ should be
$y>-1$, i.e. $F>\kappa^2/e \simeq 0.368\kappa^2$.
Before setting up a cosmological model,
the EoS $w_y$ itself as a function of $F$
is interesting.
From Eqs.(\ref{rhoy}) and (\ref{py})
one sees that the YMC exhibits an EoS of radiation
with  $p_y=\frac{1}{3}\rho_y$  and
$w=1/3$ for a large dielectric $\epsilon \gg b$
(i.e. $F\gg \kappa^2$).
On the other hand, for $\epsilon =0$ (i.e. $F= \kappa^2$),
which is called the critical point,
the YMC has an EoS
of the cosmological constant
with $p_y=-\rho_y$ and $w=-1$.
The latter case  occurs when the YMC energy density takes on
the value of the critical energy density
$\rho_y=\frac{1}{2}b\kappa^2$ \cite{Zhang07}.
It is this interesting property of the EoS of YMC,
going from $w= 1/3$ at higher energies
($F\gg \kappa^2$)
to  $w=-1$ at low energies ($F= \kappa^2$),
that makes it possible for
the scaling solution \cite{Copeland98}
for the dark energy component to exist
in our model.
More interestingly, this transition is smooth
since  $w$ is smooth function of
$y$ in the range $(-1,\, \infty)$.
Now we ask the question:
Can $w_y$ cross over $-1$?
By looking at Eq.(\ref{wy}) for $w_y$, we see that
$w_y$ only depends on the value of the condensate strength $F$.
In principle,
$w_y<-1$ can be achieved as soon as $F< \kappa^2$.
Moreover, in regards to the behavior of $w_y$
as a function of $F$,
this crossing is also smooth.
However, as shall be shown explicitly later,
when the YMC is put into a cosmological model
as the dark energy component,
together with the radiation and matter components,
to drive the expansion of the universe,
the value of $F$ can not be arbitrary,
it comes out as a function of time $t$ and
has to be determined by the dynamic evolution.
Specifically, when the YMC does not decay into
the matter and radiation,
$w_y$ can only  approaches to $-1$ asymptotically,
but will not cross over $-1$.
On the other hand,
when the YMC decays into the matter and/or  radiation,
$w_y$ does cross over $-1$,
and, depending on the strength of the coupling,
$w_y$ will settle down to an asymptotic  value $\sim -1.17$.
As a merit, in this lower region of $w_y<-1$,
all the physical quantities $\rho_y$, $p_y$, and $w_y$
behave smoothly,
there is no finite-time singularities
that are suffered by a class of
scalar models.

Now we put the YMC in to the cosmic setting,
which is assumed to be a spatially flat ($k=0$)
Robertson-Walker spacetime
\be
 ds^2= d t^2-a^2(t) \delta_{ij} dx^i dx^j . \label{metric}
 \ee
As it stands, the present universe is filled
with three kinds of
major energy components,
the dark energy,
the matter,  including both  baryons and dark matter,
and the radiation.
In our model,  the dark energy component
is represented by the YMC,
and the matter component is simply described
by a non-relativistic dust
with negligible pressure,
and the radiation component consists of CMB
and possibly other particles, such as neutrinos,
if they are massless.
Since the universe is assumed to be flat,
the sum of the fraction densities is
$\Omega=\Omega_y+\Omega_m+\Omega_r =1$,
where the fractional energy densities are
$\Omega_y= \rho_y/\rho$, $\Omega_m= \rho_m/\rho$,
and $\Omega_r= \rho_r/\rho$.
The overall expansion of the universe
is determined by the  Friedmann equations
\be \label{friedmann1}
(\frac{\dot{a}}{a})^2=\frac{8 \pi G}{3}(\rho_y+\rho_m+\rho_r),
\ee
\be \label{friedmann2}
\frac{\ddot{a}}{a}=
       -\frac{4 \pi G}{3}(\rho_y+3p_y+\rho_m+\rho_r+3p_r),
\ee
in which all these three components of energy
contribute to the source on the right-hand side of the equations.
The dynamical evolutions of the three components
are determined by their equations of of motion,
which can be written as equations of
conservation of energy  \cite{Zhang07, Zhang}:
\be \label{ymeq}
\dot{\rho}_y+3\frac{\dot{a}}{a}(\rho_y+p_y)=-Q_m-Q_r,
\ee
\be \label{meq}
\dot{\rho}_m+3\frac{\dot{a}}{a}\rho_m=Q_m,
\ee
\be \label{req}
\dot{\rho}_r+3\frac{\dot{a}}{a}(\rho_r+p_r)=Q_r,
\ee
where $Q_m$
represents the energy exchange between the YMC and the
matter, and $Q_r$ between the YMC
and the radiation,
respectively.
In the natural unit, both quantities have the dimension
of $[energy]^5$.
The couplings  $Q_m$ and $Q_r$ are phenomenological,
and their specific forms of  will
be addressed later.
The sum of Eqs. ({\ref{ymeq}}), ({\ref{meq}}), and ({\ref{req}})
guarantees  that the total energy is still conserved.
As is known, Eq.(\ref{friedmann2}) is not independent and
can be derived from
Eqs.(\ref{friedmann1}), (\ref{ymeq}), (\ref{meq}), and (\ref{req}).
It is noted that once the couplings $Q_m$ and $Q_r$
are introduced as above,
they will bring  two new parameters in our model.
When $Q_m>0$, the YMC transfers energy into the matter,
and this could be implemented, for instance,
 by the processes with
the YMC decaying into pairs of matter particles.
On the other hand, when $Q_m<0$,
the matter transfers energy into the YM condensate.
Similarly, when $Q_r>0$,
the YMC transfers energy into   the radiation.
Therefore, in the most general case of coupling,
there will be  three model parameters:
$Q_m$, $Q_r$, and $\kappa$.

In the following computations,
it is simpler to employ  the following
functions rescaled by
the critical energy density $ \frac{1}{2}b \kappa^2$
of YMC,
\be   \label{x}
x \equiv \frac{\rho_m}{  \frac{1}{2}b \kappa^2},
\ee
\be \label{r}
r\equiv \frac{\rho_r}{\frac{1}{2}b\kappa^2} ,
\ee
\be
q_m\equiv  \frac{Q_m}{\frac{1}{2}b\kappa^2},
\ee
\be
q_r\equiv  \frac{Q_r}{\frac{1}{2}b\kappa^2}  .
\ee
Here the dimensionless functions $x$ and $r$ are simply
the rescaled energy density of the matter and radiation,
respectively,
and the rescaled exchange rates $q_m$ and $q_r$
have unit of $[time]^{-1}$.
Then,  in terms of $x$, $y$, and $r$,
the dynamical evolutions given in Eqs. (\ref{friedmann1}),
 (\ref {ymeq})-(\ref{req}) can
be recast into:
\be \label{y}
\frac{dy}{d N}   +\frac{4y}{2+y}
             =-\frac{q_m+q_r }{   H h (2+y)e^y },
\ee
\be \label{x1}
\frac{dx}{d N} +3x=\frac{q_m}{  H h},
\ee
\be \label{r1}
\frac{dr}{d N} +4r = \frac{q_r}{   H h },
\ee
\be \label{hubble}
(\frac{\dot{a}}{a})^2=H^2h^2,
 \ee
where the variable $N=\ln a(t)$,
the function $h=\sqrt{(1+y)e^y+x+r}$,
and the constant $H=\sqrt{4 \pi Gb\kappa^2/3}$.
Note that $H$ is not exactly the present Hubble constant $H_0$.
From Eq.(\ref{hubble})
one can see that it is the quantity $Hh$
that determines the actual expansion rate of the universe,
and the present value of $Hh$ is identified as
the Hubble constant $H_0$.
However, as our calculations will show later,
the value of $h\sim 1.07$ at present,
so approximately $H\simeq H_0$.
It should be emphasized
that $H$ is not an independent parameter,
it is  fixed  by the model parameter $\kappa$.
Once the YMC is put into the cosmological context,
one can estimate  the order of
magnitude of  $\kappa$ as follows.
The critical density $\rho_c= 8.099h_0^2\times10^{-11}$ev$^4$
with the Hubble parameter $h_0\simeq 0.72$,
and the current value of the dark energy density
should be $\rho_y= \Omega_y\rho_c \simeq  0.7\rho_c $.
As  calculations will show,
the present value of
the factor $(1+y)e^y\simeq 0.8$ in Eq.(\ref{rhoyy}),
and $\rho_y \simeq 0.8\times \frac{1}{2}b\kappa^2 $.
So one has
\be \label{kappa}
\kappa^{1/2}  \simeq  5\times10^{-3}h_0\,\,ev.
\ee
This energy scale is much smaller than
those typical energy scales occurring
in the standard model of particle physics,
such as $\sim 10^2$ Mev for QCD,
and $\sim 10^2$ Gev for the weak-electromagnetic unification.
Therefore, for lack of an explanation
of the origin of this energy scale $\kappa$
within the standard model,
we may have to regard the YM condensate as a new physics
beyond the standard model of particle physics.
In this sense,
like other dark energy models,
the fine-tuning problem,
i.e. why $\kappa$ has just this small value,
also exists in our model.

The set of
equations   (\ref{y}) --  (\ref{hubble}) hold for a generic
stage of cosmic expansion driven by
the combination of
the radiation, the matter, and the dark energy.
In this
paper we focus on the
matter-dominated era and the subsequent accelerating era.
In particular, we like to see
how the cosmic expansion evolves
and transits from the matter-dominated
to the accelerating era.
In the remaining of the paper,
we always take the initial condition to be at the time $t_i$
of the equality of radiation-matter
(with a redshift $z= 3454^{+385}_{-392}$ \cite{spergel})
\be
\Omega_{mi}=\Omega_{ri},
\ee
with the subscript $i$ denoting the initial value.
Of course,
once the initial value $\Omega_{yi}$ for
the YMC is given,
one has immediately
\be
 \Omega_{mi}=\Omega_{ri}= \frac{1}{2}-  \frac{1}{2}\Omega_{yi}.
\ee
In order to keep the main features of the conventional
scenario of  cosmic expansion,
it is also assumed that  initially
the matter and the radiation are dominant,
$\Omega_{mi}  =   \Omega_{ri} \simeq 1/2$,
and the YMC is subdominant
\be \label{oyi}
\Omega_{yi}\ll 1/2.
\ee
In Ref.\cite{Zhao} we considered the constraints on
the YMC energy density at an earlier stage.
There the initial condition was taken
at a redshift $z\simeq 10^{10}$,
corresponding to an energy $\sim 1$ Mev,
 during the radiation stage,
when the Big Bang nucleosynthesis processes took place.
The upper bound  has been found to be
$\Omega_y\leq 0.26\, \Omega_r$ at $z\simeq 10^{10}$.
Afterwards up to $z\sim 3000$,
both $\rho_r$ and $\rho_y$ evolved
approximately in a similar way $\propto a(t)^{-3}$.
Thus, at the equality
of radiation-matter with  $z\simeq 3500$,
this consideration will give
an upper bound  $\Omega_{yi}\sim 0.1$.
Within this restriction,
the initial value $\Omega_{yi}$ may still be
allowed to vary in a very broad range.
In the following we take a safe value
\be \label{irange}
\Omega_{yi}\leq 10^{-2}
\ee
as the upper bound both for illustration purpose.
This choice has been made in concordance with the thinking
that the dark energy component
has been existing in the universe from the equality
of radiation-matter, but its initial relative contribution
is bounded by a few percent of the total,
so that during most of the history of the universe
the cosmic evolution will be
the same as in the standard Big Bang model.
Only quite recently has the dark energy component
become dominant and
modified the cosmic evolution considerably.

\begin{center}
{\em\Large 3. Non-coupling Case  }
\end{center}

The simplest case is the non-coupling with  $Q_m=0$ and  $Q_r=0$
in Eqs.(\ref{ymeq}), (\ref{meq}),  and (\ref{req}).
Then there is only one model parameter $\kappa$
that has already been  fixed  in Eq.(\ref{kappa}).
Each component evolves independently
in the expanding RW spacetime.
To solve the dynamic equations,
we take the initial YMC at the time $t_i$
to be in the broad range
\be \label{iyn}
\Omega_{yi} =  ( 10^{-10}, ~ 10^{-2}),
\ee
consistent with the restriction (\ref{irange}),
that is,  the initial value of
$\rho_{yi}$ ranges over eight orders of magnitude.
In terms of $y$, $x$, and $r$,
the initial condition above can be written as
\be \label{yi}
 y_i = (1, ~ 16.12),
\ee
\be \label{xr}
x_i =r_i =1.7\times 10^{10}.
\ee
The equations (\ref{y}) (\ref{x1}) (\ref{r1})
with $q_m=q_r=0$ are solved easily
for $\rho_y(t)$, $\rho_m(t)$, and $\rho_r(t)$,
which are shown as a function of redshift $z$
in Fig.\ref{fig1}.
Both $\rho_m(t)\propto a^{-3}(t)$
and $\rho_r(t)\propto a^{-4}(t)$
decrease monotonically at their fixed slope, respectively,
and do not level off as $t\rightarrow \infty$.
More interesting is the evolution of YMC energy density.
In the early stage
$\rho_y(t)$ is subdominant to  $\rho_m(t)$ and $\rho_r(t)$,
 and decreases at a  slope between those
 of $\rho_r(t)$ and $\rho_m(t)$,
 tracking the matter.
Later, $\rho_y(t)$ gradually levels off at
and approaches to a constant.
At $z\sim 0.35$, $\rho_y(t)$ starts to dominate
over $\rho_m(t)$ and
the accelerating expansion takes over.
This exit from the subdominant region (scaling) to
the dominant region is naturally realized.
It is important to notice that,
as long as the initial value is in the broad
range of Eq.(\ref{iyn}),
$\rho_y(t)$ always has
the same asymptotic value as $t\rightarrow \infty$.
By the way,
the first order differential equations (\ref{x1}) and (\ref{r1})
for $x(t)$ and $r(t)$
have no fixed points since $dx/dN \ne 0$ and  $dr/dN \ne 0$
during the course of evolution,
but the differential equation (\ref{y})
for $y(t)$ has a fixed point $y_f=0$ as solution of $dy/dN =0$
as $t \rightarrow \infty$.
Fig.\ref{fig2}
gives the corresponding evolution of
the fractional energy densities.
Starting with the initial value $\Omega_{ri}\simeq 1/2$,
the radiation component $\Omega_r$
has a simple evolution of monotonic decrease.
In contrast, the matter component $\Omega_m$,
starting with $\simeq 1/2$,
increases quickly and approaches to $\sim1$ around
a redshift $(1+z) \sim  174$.
At $(1+z)\sim 2.7$, $\Omega_m$
drops down and is dominated by $\Omega_y$ at $z\sim 0.35$.
The YMC component  $\Omega_y$ starts with
the very small initial value and
increases slowly and monotonically.
Around  $(1+z)\sim 2.7$, $\Omega_y$ has a quick increase,
and around  $z\sim 0.35$ it dominates over $\Omega_m$.
Observe in Fig.\ref{fig2} that the two curves of $\Omega_y$ for
the two different initial values $ 10^{-10}$ and $ 10^{-2}$,
respectively,
are almost overlapped into one curve.
This pattern of degeneracy for $\Omega_y$
demonstrates vividly the fact that
the cosmic evolution and the current status are
insensitive to the initial condition of the YMC.
As the result of evolution, at present ($z=0$),
one has $\Omega_y  \sim 0.7$, $\Omega_m  \sim 0.3$,
and $\Omega_r\sim 10^{-5}$.
Fig.\ref{fig3} shows the evolution of
 $w_y$
and that of the effective EoS $w_{eff}$ defined by
\be
w_{eff} = \frac{p_m+p_y}{\rho_m+\rho_y}.
\ee
Both approach to $-1$ as $t\rightarrow \infty$,
but,  they  can not across $-1$.
So there is no super-accelerating stage in the non-coupling model.
Looking back Eq.(\ref{wy}) and (\ref{defy}),
it is clear that in the non-coupling case
the YM field strength will always stay above
the critical value: $F\geq \kappa^2$.
The asymptotic region at $t\rightarrow \infty$
corresponds to $F=\kappa^2$.
This is a state with the dielectric constant
$\epsilon =0$.
Thus, for the non-coupling case,
no matter what kind of initial condition
is given, the YMC always settles down
to the state of $\epsilon =0$
as a result of dynamic evolution.
As for the current status
$\Omega_y  \sim 0.7$ and $\Omega_m  \sim 0.3$,
it has been achieved
for the whole range of initial values in Eq.(\ref{yi})
at the fixed model parameter
$\kappa$ in Eq.(\ref{kappa}).
Therefore, the coincidence problem is solved in this model,
but the fine-tuning problem still exists,
i.e., we do not have an answer to the question
why $\kappa$ should have
such a value as in  Eq.(\ref{kappa}).
The non-coupling case with  $Q_m=0$ and  $Q_r=0$
has also been studied in \cite{Zhang, Zhao}
with the initial condition being
taken at a redshift $z\sim 10^{10}$
in the radiation dominated stage.
There the evolution behavior found for the matter dominated era is
similar to what is obtained  here.

\begin{center}
{\em\Large 4. YMC Decaying into  Matter }
\end{center}

Consider the case that the YMC
couples to the matter component only.
In terms of the  cosmic energy densities today,
the radiation fraction  is roughly $\Omega_r \sim 10^{-5}$,
a relatively very small contribution,
much lower than the other two components.
Therefore, in this section we temporarily
neglect its coupling with the YMC
by setting $Q_r = 0 $.
There is only one free parameter $Q_m$
since $\kappa$ has been fixed.
 Then Eqs.(\ref{y}), (\ref{x1}), and (\ref{r1}) reduce to
\be \label{y2}
\frac{dy}{d N} + \frac{4y}{2+y}=-\frac{q_m}{  H h (2+y) e^y},
\ee
\be \label{m2}
\frac{dx}{d N}+ 3x=\frac{q_m}{  H h},
\ee
\be \label{r2}
\frac{dr}{d N}+4r=0.
\ee
The dynamic evolution of the radiation is independent
of the other two,
and $\rho_r(t) \propto a^{-4}(t)$,
as is seen in Eq.(\ref{r2}).
To proceed further,
one needs to know the coupling $q_m$ to
solve  Eqs. (\ref{y2}) and (\ref{m2}).
As mentioned earlier,
in general there are  two kinds of models
depending on whether $Q_m>0$ or  $Q_m<0$.
In in section, we examine the model $Q_m>0$ with the YMC
decaying constantly into the matter.
We call this the Model 1.
It can be generically expressed as
\be \label{qm}
Q_m =\Gamma \rho_y,
\ee
where  $\Gamma >0$ is of dimension $[time]^{-1}$
and measures the decay rate of the YMC energy density
 as well as the production rate of the  matter energy density.
Here $\Gamma$ is simply taken as
a model parameter describing
phenomenologically the interactions
between the YMC and the matter.
In the following we discuss several cases
for the parameter $\Gamma$.
Substituting Eq.(\ref{qm})
into Eqs. (\ref{y2}) and (\ref{m2}) yields
\be \label{xx}
\frac{dx}{d N}=\frac{\Gamma}{H}\frac{(1+y)e^y}{h}-3x,
\ee
\be \label{yy}
\frac{dy}{d N}=-\frac{\Gamma}{H}\frac{1+y}{(2+y) h}-\frac{4y}{2+y}.
\ee

1. Consider the simple case of a constant rate with
\be \label{gamma}
\Gamma/H=0.5.
\ee
We have taken this magnitude of the rate $\Gamma$,
so that the present status of the universe
will be  $\Omega_y\simeq 0.7$
and $\Omega_m\simeq 0.3$ as  the outcome from our computation.
Interestingly,
in order to achieve this status,
as a model parameter,
the decay rate of the YMC
needs to be of the same order of magnitude as
the expansion rate of the universe,
i.e. $\Gamma \sim  H$.
Now at the time $t_i$ with $z\simeq 3454$
the initial YMC energy density
is taken to be in the range
\be \label{iy}
\Omega_{yi} =  ( 10^{-10}, ~~ 3\times 10^{-3}),
\ee
similar to Eq.(\ref{iyn}),
which corresponds to
\be \label{yi2}
y_i = (1, \, 15).
\ee
The initial values for the  matter and
the radiation components are given by
\be \label{xri}
x_i =r_i = 1.0\times 10^{10},
\ee
which are a little bit smaller than
 that in Eq.(\ref{xr}).
This is because in the case here the matter are being generated
out of the decaying YMC during the course  of evolution.
Consequently,
smaller initial values $x_i$ and $r_i$
are needed to arrive at the current status.

Given the different initial values of $\Omega_{yi}$ in Eq.(\ref{iy}),
the corresponding initial values of
$\Omega_{mi}=\Omega_{ri}=(1-\Omega_{yi})/2$
also vary by a small amount.
But for the different values of $y_i$ in Eq.(\ref{yi2})
we have taken the same set of values in Eq.(\ref{xri})
since the initial total energy density
$\rho_{yi}+ \rho_{mi}+\rho_{ri}$ itself
at $z\simeq 3454$ has some errors.
The results are given as functions of the redshift $z$
in Fig.\ref{fig4}, Fig.\ref{fig5}, and Fig.\ref{fig6} .
The evolution of $\rho_y(t)$ is also similar to the
non-coupling case. During the early stage $\rho_y(t)$ is lower than,
and keeps track of $\rho_m(t)$.
Later, $\rho_y(t)$ levels off and
approaches to a constant, and around $z\sim 0.48$, it starts to
dominate over $\rho_m(t)$,
 and the accelerating stage begins.
In fact in Fig.\ref{fig4} for three different $y_i$
there are three curves of the matter $\rho_m(t)$, respectively.
However, these three curves are too close to each other
so that they are overlapped.
The similar is for $\rho_r(t)$.
Note that, in the presence of coupling $Q_m$,
the evolution of $\rho_m(t)$ is different from the non-coupling case,
in that at the late stage around $z\sim 0$
 it also levels off just like  $\rho_y(t)$.
In this sense, the evolution of the matter component is
sort of bound to the YMC.
In fact, as $t\rightarrow \infty$,
the set of Eqs.(\ref{xx}) and (\ref{yy}) have
the asymptotic behavior
\be \label{asym}
\frac{dx}{dN}  \rightarrow 0 ,  ~~ ~~ \frac{dy}{dN}
          \rightarrow 0,
\ee
and both $\rho_y(t)$ and $\rho_m(t)$ have asymptotic
values.
This  will be addressed later in section 6.

As a novel feature of the coupling model,
in contrast to the non-coupling case,
now the EoS of the YMC  as a function of time $t$,
$w_y(t)$   crosses over $-1$ around $z\sim 2$,
takes  a value $w_y\simeq -1.1$ at present $z=0$,
and approaches $w_y \simeq -1.17$ asymptotically,
as shown in Fig.\ref{fig6}.
This occurrence of crossing over $-1$
in this model can be understood,
since the coupling makes the YMC to
loss energy into the matter,
consequently,  the YM field strength $F$
will drop down below the critical value $\kappa^2$,
leading to $\epsilon =by <0$ and $w_y<-1$ as in Eq.(\ref{wy}).
This can also be arrived by looking at
the asymptotic region determined by the equation
$\frac{dy}{dN}= 0 $,
which by Eq.(\ref{yy}) is just
\be \label{asymy}
\frac{\Gamma}{H}  \frac{ 1+y}{h} +4y=0.
\ee
Recall that for the non-coupling $\Gamma =0$
the asymptotic value is $y_f=0$, yielding $w_y =- 1$
by Eq.(\ref{wy}).
Once $\Gamma >0$, Eq.(\ref{asymy}) yields
an asymptotic value $y_f<0$ as the solution,
 hence $w_y<-1$.
Thus,  when transferring energy to the matter,
 the YMC  will eventually settle down in the state
of  $w_y < -1$,
which is equivalent to a negative dielectric $\epsilon <0$.
Recently, there are some observational indications
that the current value of EoS of dark energy
$w$ is less than $-1$,
for instance,
$w=-1.023\pm{0.090}(stat)\pm{0.054}(sys)$
from the 71 high redshift
supernovae  discovered during the first year SNLS \cite{Astier},
and $w =-1.21^{+0.15}_{-0.12}$ from the blind analysis
of 21 high redshift supernovae by CMAGIC technique  \cite{Knop}.
Of course, this is still to be observationally
examined with higher confidence level in future.
However, the crossing over $-1 $ would
be difficult for scalar models,
except for quintom models at a price of
introducing two scalar fields
and an artificially designed potential \cite{Zhao06}.
As we just have demonstrated,
in the YMC dark energy model with coupling $Q_m>0$,
this crossing is realized naturally.
In general, the asymptotic value of $w_y$ at $t\rightarrow \infty$
and the current value $w_y$ at $z=0$ as well,
are determined by the asymptotic value of $y$ through Eq.(\ref{wy}),
and the latter is obtained from
the combination of Eqs.(\ref{yy}) and (\ref{asym}),
and thus depends on the ratio  $\Gamma/H$.
Thus the asymptotic value of $w_y$
is determined by the ratio $\Gamma/H$
of the two parameters of our model.
For instance, the observed Eos of dark energy
$w=-1.023$  from SNLS \cite{Astier} can be obtained,
in our model,
 by taking a little smaller decay rate $\Gamma/H= 0.13$ and
slightly higher initial densities $x_i=r_i= 1.5\times 10 ^{10}$,
and the value $w=-1.21$ from the analysis by CMAGIC \cite{Knop}
can be obtained by taking
a bit larger decay rate $\Gamma/H= 0.81$ and
a slightly lower densities $x_i=r_i= 0.7\times 10 ^{10}$,
respectively.
Fig.\ref{fig6} also shows that the
effective $w_{eff}$ can not cross $-1$ yet,
and its asymptotic value is $\sim -0.96$.
Interestingly,  this model predicts
that in the upcoming future
the dark energy density $\rho_y $ will remain a constant
slightly larger than today,
and that the matter energy density  $\rho_m $ will be a constant
slightly lower than  today.
Eventually the universe will settle down to a steady state with
$\Omega_y\sim 0.85$, $\Omega_m \sim 0.15$, and $\Omega_r \sim 0$.

Therefore, this model yields a
picture of evolutional cosmos,
the early part of which can account for
the the past history of the expanding universe,
i.e., that of the standard Big-Bang model,
and the late part of which, i.e. the  future of the universe,
is similar to the Steady State Model
\cite{bondi-gold} \cite{hoyle}.
As a plus for the YMC model,
there are no Big Rip singularities in finite time,
since all the quantities $\rho_y$, $p_y$, and $w_y$
are smooth function of the time $t$.

Although in the above we have only presented the results
for the initial values
within the range $\Omega_{yi}> 10^{-10}$ given in Eq.(\ref{iyn}),
as a matter of fact,
we have worked out for even lower values $\Omega_{yi} <10^{-10}$.
For instance, we have calculated
the case of $\Omega_{yi} \simeq 2\times  10^{-12}$, i.e., $y_i= -0.9$.
the initial energy density $\rho_{yi}$ are even lower than
the current values  $\rho_{y} \simeq 0.7 \rho_c$.
The details are given by the curves denoted by
 $y_i = -0.9$ in Fig.\ref{fig4}, Fig.\ref{fig5}, and Fig.\ref{fig6}.
The evolution is such that
$\rho_{y}(t)$ starts initially
from the given very low value,
increases (instead of decreasing) very quickly with time $t$ ,
and approaches its corresponding asymptotic value
$\simeq 0.7\rho_{c}$, as is seen in Fig.\ref{fig4}.
The evolutions for $\rho_m(t)$ and $\rho_r(t)$
are similar to the cases in Eq.(\ref{iy}).
Thus,
except the initial increasing  of $\rho_y(t)$,
the current status of the universe
is the same as those in the range of Eq.(\ref{iy}).

2. Now consider a case of the YMC decaying
into fermion pairs.
As is known in QED, a constant electric field
is unstable against decay into pair
of particles \cite{Schwinger}.
Analogously,  a constant `electric'  $SU(3)$ Yang-Mills
 field configuration of QCD
is unstable against decay into quark pairs
and gluon pairs \cite{Matinyan}
\cite{Gyulassy} \cite{Kajantie}.
Calculations have shown that
for a gauge group $SU(N)$,
due to the decay into fermions pairs and  gauge boson pairs,
the average local color electric field decreases
at a rate
\be \label{decayrate}
\frac{\partial E}{\partial t}  =-cE^{3/2},
\ee
 where c is a dimensionless
 constant \cite{Gyulassy} \cite{Kajantie} \cite{Zhang07}.
 Notice that this rate is derived from
 the energy-momentum conversation of pair creation.
From the point of view of particle physics,
the $SU(2)$ YM field in our case should be allowed to
have some fundamental interactions
with other microscopic particles.
At the moment we do not intend to proceed further to build up
the detail of its interaction.
Instead, we simply assume that,
similar to the $SU(3)$ gauge field in QCD,
the YM condensate,  here a constant in space,  also has the property
of being unstable against decay into other particles,
and the decay rate of the field strength has the same form as
  Eq.(\ref{decayrate}).
Consequently, this in turn will give rise to
the decay rate of the YMC energy density
\be \label{energydecay}
\frac{ \partial \rho_y}{\partial t} = -\Gamma\rho_y.
 \ee
Making use of the chain rule relation
$\frac{ \partial \rho_y}{\partial t}
   = \frac{ \partial \rho_y}{\partial E}
     \frac{ \partial E}{\partial t} $
   and Eq.(\ref{decayrate}),
one obtains the following expression for
the decay rate \cite{Zhang07}
\be \label{decayintopair}
\Gamma=2c\kappa^{\frac{1}{2}}
 \frac{2+y}{1+y}e^{\frac{y}{4}}.
\ee
depending on the coefficient constant $c$,
which is treated as
the model parameter in place of  $\Gamma$.
We choose the value of  $c=0.125 H/\kappa^{1/2}$,
so that
\be \label{decay2}
\frac{\Gamma}{H}=0.25 \frac{2+y}{1+y}e^{\frac{y}{4}}.
\ee
As our computations show,
the value of  variable $y$
at present stage is very small $y\sim 0$,
so Eq.(\ref{decay2}) yields $\Gamma\sim 0.5 H$,
quite close to that in Eq.(\ref{gamma}).
The initial condition is taken to be
the same as Eqs.(\ref{yi2}) and (\ref{xri}).
Figs  \ref{fig7}, \ref{fig8}, and \ref{fig9} show the results,
which are very similar to that in the previous case
of the constant rate.
From Eq.(\ref{meq}) it is seen that,
in the late-time asymptotic region when $\dot \rho_m \simeq 0$,
the matter generation rate is estimated as
 $Q_m \simeq 3H\rho_m\simeq 10^{-46}g\, cm^{-3}s^{-1}$,
a very small rate,
equivalent to generation of $\sim 0.3$ protons in
a cubic kilometer per year.
This value is approximately equal to
that in the Steady State Model
 \cite{bondi-gold} \cite{hoyle}.
Thus in our model
the particle pairs are continuously generated,
at a very low rate,
out of the vacuum filled with the YMC.

As is known, in order to continuously generate the cosmic matter,
the Steady State Model has to
introduce some $C$-field with negative energy \cite{hoyle},
which is problematic in a physical theory.
Here in our model,
it is the effective quantum YMC
plays the role of a matter-generator,
there is no negative energy to occur in the proper range.
The YMC has a positive energy
and a negative pressure.

From Eq.(\ref{gamma}) and (\ref{decay2})
it seems that the overall behavior of
the dynamic evolution is not sensitive
to the particular form of the coupling $\Gamma$,
as long as its magnitude is $\Gamma\sim 0.5H$.
This has been confirmed in our examinations.
For example, we have also investigated
the case with the decay rate of the  form
\be \label{cy}
\Gamma/H=0.5e^{-y},
\ee
and the results are very similar to that in the previous case.
Therefore, in Model 1 of the YM-matter coupling
with the coupling $Q_m= \Gamma \rho_y$,
as long as $\Gamma$ is constant or depends on $\rho_y$,
the overall features of the dynamic evolution
are similar.
We have also studied
the cases with $\Gamma$ depending on the matter $\rho_m$.
It is found that,
if the decay rate $\Gamma$ depends on the matter
  $\rho_m$,
for instance
\be \label{by}
 \Gamma/H= bx,
\ee
where $b$ is some constant and $x$ is defined in Eq.(\ref{x}),
then for $ b \leq 10^{-5}$
the evolution will be similar to the non-coupling case.
When the constant $b\geq 10^{-3}$,
the decay rate  is too fast, and
at start $\rho_y(t)$ drops down quickly, later it
increases to its asymptotic value from below.
To keep the paper short,
we do not demonstrate these detailed graphs
of the cases of Eqs.(\ref{cy}) and (\ref{by}).

\begin{center}
{\em\Large 5.  Matter Decaying into YMC }
\end{center}

In section
we study the situation in which the matter decaying constantly
into the YMC with  $Q_m<0$,
just opposite to the Model 1.
We call this the Model 2.
It can be generically expressed as
\be
Q_m =-\Gamma\rho_m,
\ee
i.e. the matter transports energy into the YM condensate.
Then  Eqs. (\ref{y2}) and (\ref{m2}) reduce to
\be
\frac{dx}{d N}  + 3x=-\frac{\Gamma}{H}\frac{x}{h},
\ee
\be
\frac{dy}{d N}+\frac{4y}{2+y}=\frac{\Gamma}{H}\frac{x}{(2+y)e^y h}.
\ee

1. Consider the simple case of a constant decay rate
\be \label{0.02}
\Gamma/H= 0.02.
\ee
Again,  here the value $0.02$ has been  taken,
so that the resulting energy densities from our computation
will be $\Omega_y\simeq 0.7$ and $\Omega_m\simeq 0.3$ at present.
So in the Model 2
the decay rate of the matter into the YM condensate
needs to be almost two order of magnitude smaller than the expansion rate.
The initial values of $x_i$ are taken to be
\be \label{-xri}
x_i=r_i\simeq 1.8\times10^{10}.
\ee
Since the matter is decaying and constantly being converted
into the YMC component,
a slightly larger initial value of the matter has been taken
than that in Eq.(\ref{xri}).
The initial value of the YMC
is taken to be $y_i = (1,15)$, the same as Eq.(\ref{yi2}).
The results are given in Figs.\ref{fig10}, \ref{fig11}, \ref{fig12}.
Now $w_y$ does not cross over $-1$.
It is interesting to find out that
the evolution of the YM condensate
behaves differently for two different ranges of $y_i$.
For the higher range
\be \label{y5}
y_i = (5,~15),
\ee
corresponding to
$\Omega_{yi} = ( 5\times 10^{-8},~~ 3\times 10^{-3})$,
all the quantities have  similar evolution as
 the non-coupling case in Figs. \ref{fig1}, \ref{fig2}, \ref{fig3}.
For the lower range
\be \label{y1}
y_i= (1,~5),
\ee
however,
the YMC has an instantly sudden increase
during the initial stage,
and quickly catches up the evolution pattern of the $y_i=5$ case.
This is in contrast to the smooth behavior on the higher range.
Thus, in order to have a rather smooth evolution
within the Model 2,
the initial value $y_i$ should
be given by the higher range in Eq.(\ref{y5}).
Moreover,
we have also found that the overall behavior
of the dynamic evolution is not sensitive to
the particular form of the coupling $Q_m$.
For instance, we have checked a case of
\be \label{0.02x}
\Gamma/H=0.02e^{-x},
\ee
and the resulting evolutions are similar
to the case of Eq.(\ref{0.02}).
Thus, the coincidence problem can also be solved in Model 2.

So far in Model 1 and Model 2,
in regards to the coupling between
the YMC and the the matter,
we  have not explicitly distinguished
the baryons and the dark matter,
and have assumed, for simplicity,
the same coupling $Q_m$
for both the baryons and the dark matter.
We can roughly estimate the current value of
the cross section
 corresponding to the collisions
involving the baryons
as in Eqs.(\ref{gamma}), (\ref{decay2}), (\ref{cy}),
(\ref{by}) with $b\leq 10^{-5}$,
(\ref{0.02}), and (\ref{0.02x}).
For instance, take the baryon decay rate
$\Gamma \sim  bx H$ as in Eq.(\ref{by}).
Then,  by definition,  the rate is $\Gamma\sim v n\sigma$,
where $v$ is the baryon velocity,
the baryon number density
is $n=\rho_b/m_b\sim 0.04 \rho_c/m_b$,
and $\sigma$ is the crossing section
for the collisions between the baryons and the YM gauge bosons
for this type of interaction .
Then we can get an estimate:
\be
\sigma \sim 25 b m_b H/v \rho_c.
\ee
Taking $b\sim 10^{-5}$, $v\sim 10^{3}$km/s,
 the baryon mass $m_b\sim 0.94$ Gev,
the current Hubble constant for $H$,
one has $\sigma \sim 6\times 10^{-26}$ cm$^{2}$.
This is an order lower than the Thomson's
cross section $\sigma_T\simeq 6.7\times 10^{-25}$ cm$^{2}$
for in QED.
Similarly, letting  $\Gamma \sim  0.5 H$ as
in Eqs.(\ref{gamma}) and  (\ref{decay2})
for the YMC decaying into baryons,
we would get $\sigma \sim 20\times 10^{-25}$ cm$^{2}$
analogously,
slightly greater than $\sigma_T$.
Therefore, given this magnitude for
the cross section $\sigma$
in both cases,
we would, in principle,
be able to observe this kind of
interactions occurring,
either with the baryon being decaying into the YM boson pairs
or the baryon pairs jumping out of the vacuum.
However, as said earlier,
the rate $\Gamma $ for this kind of events is too low,
giving $Q_m  \sim 0.3$ proton generated
in one cubic kilometer per year.
For the Galaxy of a volume $\sim 10^3$(kpc)$^3$,
this rate is roughly equivalent to
an amount of mass $\sim  10^{-6}M_{\odot}$
generated per year, a small production rate.
The chance may be small for  directly detecting the event.
Even if future experiments rule out or restrict
the coupling with the baryons,
one has to drop it or reduce its magnitude as a model parameter.
Nevertheless, the coupling with the dark matter
probably still remains.
This is because
the dark matter is usually assumed not to have interactions
with ordinary particles,
such as baryons, photons, etc.
So it is difficult to directly detect
productions of dark particle pairs and
decays of dark particles.

\begin{center}
{\em\Large 6. Coupling with Both Matter and Radiation }
\end{center}

It is quite natural  to allow the  YMC to couple with
both the matter and the radiation  simultaneously.
Now we study Model 3 that the YMC decays
into the matter and the radiation as well:
\be
Q_m=\Gamma\rho_y>0, \,\,\,\,\,\, ~~Q_r=\Gamma'\rho_y>0.
\ee
Then Eqs. (\ref{y}) -- (\ref{r1}) reduce to
\be \label{yyy}
\frac{dy}{d N}=
  -\frac{\Gamma+\Gamma'}{H}\frac{1+y}{(2+y) h}-\frac{4y}{2+y},
\ee
\be \label{xxx}
\frac{dx}{d N}=\frac{\Gamma}{H}\frac{(1+y)e^y}{h}-3x,
\ee
\be \label{rrr}
\frac{dr}{d N}=\frac{\Gamma'}{H}\frac{(1+y)e^y}{h}-4r.
\ee

Consider the case of the constant decay rates
\be \label{rate}
\Gamma/H=0.5, ~~~   \Gamma'/H=1.8\times 10^{-4}.
\ee
Note that $\Gamma'$ is lower than $\Gamma $ by
three orders of magnitude.
These values
of coupling are taken so that the current values are
$\Omega_y\simeq0.7$, $\Omega_m\simeq0.3$,
$\Omega_{\gamma} \simeq
8.6\times 10^{-5}$ (including massless neutrinos).
The initial
condition is the same as in Eqs.(\ref{yi2}) and (\ref{xri}).
The results are given in
Figs.\ref{fig13}, \ref{fig14}, \ref{fig15}.
As before, the particular form of the couplings is not important
for the overall behavior of evolution.
For instance, we have also examined the case
\be
\Gamma/H=0.25 \frac{2+y}{1+y}e^{\frac{y}{4}}, ~~~~
\Gamma'/H=0.9\times 10^{-4} \frac{2+y}{1+y} e^{\frac{y}{4}} ,
\ee
based on an analogous consideration to Eq.(\ref{decay2}).
The  evolution is similar to the case of Eq.(\ref{rate}).
In  these two  cases of Model 3,
due to the couplings with the YMC,
both the energy densities, $\rho_m$ and $\rho_r$,
level off around $z\sim 0$,
and $w_y$ crosses over $-1$ around $z\simeq 2.5$.
Thus all the three components of cosmic energy
will remain constant in future,
and the state of the universe will keep almost
as it is today.
This is similar to
the Steady State universe \cite{bondi-gold} \cite{hoyle}.
Hence,
according to Model 3,
the past history of the universe is
consistent with the conventional standard Big Bang model,
and from now on,  the cosmic evolution  tends to that of
a steady state,
that is, the universe will remain similar to it is today,
but with $\Omega_y\sim 0.85$, $\Omega_m\sim 0.15$,
and $\Omega_r\sim 10^{-5}$.

We have also studied the case with both
the matter and radiation decaying into the YMC.
We call this Model 4.
The couplings are such that
\be
Q_m= -\Gamma \rho_m<0, ~~~\,\,
Q_r =- \Gamma' \rho_r<0.
\ee
We take
\be
\Gamma/H= 0.02, ~~~ \Gamma'/H =1.8\times 10^{-4} .
\ee
The initial condition for the YMC is the same as
Eqs.(\ref{y5}) and (\ref{y1})
and the initial densities for the matter and radiation
are the same as in Eq.(\ref{-xri}).
The resulting evolution
is qualitatively similar to those in the Model 2 with $Q_m<0$.
Approximately,  $\rho_m(t)\propto a(t)^{-3}$,
$\rho_r(t)  \propto a(t)^{-4} $,
and $\rho_y(t)$ also has a scaling solution
and exits the scaling regime,
just like in the non-coupling case.
The EoS of the YMC $w_y$ approaches to $-1$
from above,
but does not cross over $-1$.
To keep the paper short and concise,
we will not repeat these details and not
give the corresponding graphs here anymore.

\begin{center}
{\em\Large 7. Asymptotic Behavior and Stable
            Attractor}
\end{center}

It is interesting to investigate the  asymptotic behavior
of the dynamical evolution.
First we study the Model 1 with the
YMC coupling to the matter component only.
Now since the evolution of the the radiation
component is independent of
the the YMC and the matter,
and the value of $r$  at late time is much less
than the other variables,
so it can be neglected in the analysis of the  fixed point.
To find the fixed points,
one sets $dx/dN=dy/dN=0$ in  Eqs. (\ref{y2}) and (\ref{m2}),
and obtains the relations at the fixed point:
\be \label{fixed1}
x_f=-\frac{4}{3}y_f e^{y_f},
\ee
\be \label{fixed2}
\frac{\Gamma}{H}(1+\frac{1}{y_f})
    = -4\sqrt{1-\frac{y_f}{3}}{e^{y_f/2}},
\ee
where the sub-index $f$ refers to the the respective values at
the fixed point.
From these two equations one can write
the asymptotic value of the fractional matter density
\be
\Omega_{mf}= \frac{4y_f}{y_f-3}.
\ee
Eqs. (\ref{fixed1}) and (\ref{fixed2})
depend on the value of the ratio $\Gamma/ H$,
so does the solution $(x_f,\,  y_f)$.
Here we consider the case that $\Gamma/ H$ is constant.
Because $\Omega_{mf} $ must be larger than 0 and smaller than 1,
so $y_f$ must be in the range from $-1$ to $0$  at the fixed point.
Thus,  as long as $\Gamma/H \in(0,\infty)$
there will exist fixed points.
To be specific, consider  $\Gamma/H=0.5$.
In the region of physics there is only one fixed point:
\be \label{fixedpoint}
(x_f, \, y_f) = ( 0.13666, \,  -0.11499 ).
\ee
This is, in terms of the respective densities,
\be
(\rho_{mf}, \, \rho_{yf})
= \frac{1}{2}b\kappa^2( 0.13666  \,  , 0.78888   ).
\ee
The stability of this fixed point can be analyzed
in the conventional way as follows.
Because the two equations for $x$ and $y$ are nonlinear,
a local analysis can be given
by linearizing the two evolution Eqs.
(\ref{xx}) and (\ref{yy}).
By a standard procedure,
expanding  $x=x_f+\varepsilon$ and
$y=y_f+\eta$, where $\varepsilon$ and $\eta$
are small perturbations around the fixed point,
and keeping up to the first order of small perturbations,
Eqs. (\ref{xx}) and (\ref{yy}) reduce  to
\ba
\frac{d}{d N}
\left(
 \begin{array}{c}
 \varepsilon\\
 \eta
  \end{array}
 \right)
 = M
 \left(
 \begin{array}{c}
 \varepsilon\\
 \eta
  \end{array}
 \right),
\ea
where $M$ is a $2\times 2$ matrix depending on the values of
$x_f$, $y_f$, and $\Gamma/H$, whose the elements are
\[
M_{11}
=-\left[\frac{\Gamma}{H}\frac{(1+y_f)e^{y_f}}{2h_f^3}+3\right],
\]
\[
M_{12}=\frac{\Gamma}{H}\frac{e^{y_f}}{h_f}\left\{(1+y_f)[1-\frac{(2+y_f)
        e^{y_f}}{2h_f^2}]+1 \right\},
\]
\[
M_{21} =\frac{\Gamma}{H}\frac{1+y_f}{2(2+y_f)h_f^3},
\]
\[
M_{22} =\frac{\Gamma}{H}
    \left[\frac{(1+y_f)e^{y_f}}{2h_f^3}
          -\frac{1}{(2+y_f)^2h_f}\right]-\frac{8}{(2+y_f)^2},
\]
where $h_f=\sqrt{(1+y_f)e^{y_f}+x_f+r_f}$.
 The general solution
for the linear perturbations is of the form \be
\varepsilon=C_1e^{\mu_1 N}+C_2e^{\mu_2 N} \ee \be \eta=C_3e^{\mu_1
N}+C_4e^{\mu_2 N} \ee where  $\mu_1$ and $\mu_2$ are eigenvalues
of $M$. If they are both negative, the fixed point $(x_f, y_f)$ is
stable, and the solution is called an  attractor. For the case of
$\Gamma/H=0.5$ one finds  the matrix \ba M= \left(
\begin{array}{cc}
-3.22146 &  0.50112 \nonumber\\
 0.13180 &  -2.17625
\end{array}
\right), \ea and its two eigenvalues  $\mu_1=-3.28123$ and
$\mu_2=-2.11648$, respectively, both  negative.
Thus  the fixed
point of this model is stable, and is an attractor.
For illustration,
we plot in Fig.\ref{fig16} the
phase graph of trajectories, each trajectory starts with a
different initial condition, and ends up at the fixed point of
Eq.(\ref{fixedpoint}).
As a matter of fact,
one can check that the asymptotic behavior of
other cases of Model 1 are also
stable  fixed points.

The analysis of the asymptotic behavior
can be done analogously
for Model 3 with the couplings to the matter and the radiation.
Setting $dx/dN =dy/dN = dr/dN =0$ in
Eqs. (\ref{yyy}), (\ref{xxx}) and (\ref{rrr})
one has the following three relations
at the fixed point:
\be \label{yf}
4y_f=-\frac{\Gamma+\Gamma'}{H}\frac{1+y_f}{ h_f},
\ee
\be \label{xf}
3x_f=\frac{\Gamma}{H}\frac{(1+y_f)e^{y_f}}{h_f},
\ee
\be \label{gammaf}
4r_f =\frac{\Gamma'}{H}\frac{(1+y_f)e^{y_f}}{h_f}.
\ee
The fixed point $(x_f, y_f, r_f)$ as the solution
of this set of equations
depends on the ratios of
rates  $\Gamma/H$  and $\Gamma'/H $ as well.
Consider the constant
$\Gamma$ and  $ \Gamma'$.
The local analysis of the stability of the fixed point can be
carried out similarly.
By setting  $x=x_f+\varepsilon$,
$y=y_f+\eta$, and $r=r_f+\gamma$,
where $\varepsilon$, $\eta$, and $\gamma$
are small perturbations around the fixed point,
one has the equations
\ba
\frac{d}{d N}
\left(
 \begin{array}{c}
 \varepsilon\\
 \eta \\
 \gamma
  \end{array}
 \right)
 = M'
 \left(
 \begin{array}{c}
 \varepsilon\\
 \eta\\
 \gamma
  \end{array}
 \right),
\ea
where $M'$ is a $3\times 3$ matrix depending on the values of
$x_f$, $y_f$, $\gamma_f$,  $\Gamma/H$ and $\Gamma'/H$,
whose the elements are:
\[
M'_{11}= -\left[\frac{\Gamma}{H}\frac{(1+y_f)e^{y_f}}{2h_f^3}+3\right],
\]
\[
M'_{12} =\frac{\Gamma}{H}\frac{e^{y_f}}{h_f}\left\{(1+y_f)
         [1-\frac{(2+y_f)e^{y_f}}{2h_f^2}] +1 \right\},
\]
\[
M'_{13} = -\frac{\Gamma}{H}\frac{(1+y_f)e^{y_f}}{2h_f^3},
\]
\[
M'_{21} =\frac{\Gamma+\Gamma'}{H}\frac{1+y_f}{2(2+y_f)h_f^3},
\]
\[
M'_{22} = \frac{\Gamma+\Gamma'}{H}
       \left[\frac{(1+y_f)e^{y_f}}{2h_f^3}
                   -\frac{1}{(2+y_f)^2 h_f}\right]
          -\frac{8}{(2+y_f)^2} ,
\]
\[
M'_{23} = \frac{\Gamma+\Gamma'}{H}\frac{1+y_f}{2(2+y_f)h_f^3},
\]
\[
M'_{31} =-\frac{\Gamma'}{H}\frac{(1+y_f)e^{y_f}}{2h_f^3},
\]
\[
M'_{32} = -\frac{\Gamma'}{H }\frac{(2+y_f)e^{y_f}}{h_f}
          \left[\frac{(1+y_f)e^{y_f}}{2h_f^2}
          +1  \right]
\]
\[
M'_{33} = -\left[\frac{\Gamma'}{H}\frac{(1+y_f)e^{y_f}}{2h_f^3}+4\right].
\]
Consider the specific case
$\Gamma/H=0.5$ and $ \Gamma'/H=1.8\times 10^{-4}$.
Substituting these into
Eqs.(\ref{yf}),  (\ref{xf}) and (\ref{gammaf})
yields the unique fixed point given by
\be \label{3fp}
(x_f, ~ y_f,  ~ r_f) = (0.13666,    ~ -0.11503,
               ~  4\times10^{-5}),
\ee
and the matrix
\ba
M'=
\left(
\begin{array}{ccc}
-3.22149 & 0.50110 & -0.22149 \nonumber\\
0.13187 & -2.17631 & 0.13187 \nonumber\\
-0.00008 & 0.00045 & -4.00008
\end{array}
\right).
\ea
The three eigenvalues of the matrix $M'$ are found to be
 $-4.00015$, $-3.28124$ and $-2.11649$,
each being  negative.
Therefore,  this attractor of the Model 3 is also stable.
Notice that there are three quantities  $(x,y,r)$
 in the Model 3,
so we need the two phase
graphs of trajectories.
They are plotted in Fig.\ref{fig17} for  $(x,y)$
and Fig.\ref{fig18} for $(r,y)$, separately.
Each trajectory starts with a different initial condition,
and ends up at the fixed point of  Eq.(\ref{3fp}).
One can check the dynamics of
other cases in Model 3 also have a stable attractor.

As for Model 2 and Model 4,
only $\rho_{y}(t)$ has an asymptotic constant value
and has a stable attractor.

\begin{center}
{\em\Large 8. Conclusion and Discussion.}
\end{center}

Our motivation of this study
is to investigate the coincidence problem
for the cosmic dark energy in a spatially flat universe.
We have presented a detailed and comprehensive analysis of the model
of the effective YMC dark energy
interacting with the  matter and  radiation.
This work has been an extended development
of our previous work on the non-coupling YMC dark energy model.
Through the Friedmann equation and
the dynamic equations for each cosmic component,
once the couplings between these components are specified,
the overall cosmic evolution is fully determined
by the initial conditions of these three components.
We have studied the evolution for the matter-dominated era
starting from the
equality of radiation-matter at $z\sim 3454$.

The major results of this work are the following.

Given the initial dominant matter and radiation
$\Omega_{mi}=\Omega_{yi}\simeq 1/2$ and
the subdominant YMC energy density $\Omega_{yi}\leq 10^{-2}$,
no matter what kind of coupling between
the YMC and the matter,
or between the YMC and the radiation,
the evolution is such that the YMC is subdominant to,
and keeps track of,  the matter,
until later,  at a redshift $z\sim  0.48$ for a coupling $Q_m>0$,
or $z\sim 0.35$ for a coupling $Q_m<0$,
the YMC becomes dominant over the matter.
The era is followed by a subsequent accelerating era driven by
the dominant YM dark energy.
As the evolution outcome, the universe arrives at
the present state  with $\Omega_y\sim 0.7$,
$\Omega_r\sim 0.3$, and $\Omega_r\sim 10^{-5}$.
It is very important to note that
this has been achieved
for a variety of coupling forms $Q_m$ and $Q_r$,
and, nevertheless,
under  a very broad range
of initial condition
$\Omega_{yi}\simeq (10^{-10}, \, 3\times10^{-3})$.

If the YMC decays only into the matter,
as a result of the coupling in Model 1,
to achieve the present state of the universe,
the decay rate  needs to be  of the same order of magnitude
of the expansion rate of the universe,
i.e.,  $\Gamma\sim 0.5H $.
Moreover, for  this coupled system,
as $t \rightarrow \infty $,
both $\rho_y$ and $\rho_m$ asymptotically approach to
constants, respectively.
That is, for the system there is a unique attractor.
Furthermore, as our analysis has shown,
this attractor is stable.
As an interesting behavior,
the EoS for the YMC  $w_y$  always
crosses over $-1$ around $z\simeq 2.5$,
and the present value is $w_y\sim -1.1$.
This crossing $-1$ seems to be favored by
the recent preliminary observations on SN Ia.

When the YMC decays into
both the matter and the radiation as in Model 3,
there are two parameters $\Gamma\sim 0.5 H$
and $\Gamma'\sim 1.8\times 10^{-4} H$,
representing the respective decay rate.
The evolutional behavior is almost the same as Model 1,
and $w_y$ crosses over $-1$.
Moreover, the radiation energy density also
asymptotically approaches to a constant,
as $t\rightarrow \infty$,
like the YMC and the matter components.
Most of the conclusions
are the same as of Model 1.

On the other hand,
if the matter decays into the YMC as in Model 2
with a rate $\Gamma \sim 0.02 H$,
or if both the matter and radiation decay into the YMC
as in Model 4 with rates $\Gamma \sim0.02 H$
and $\Gamma' \sim 1.8\times 10^{-4} H$, respectively,
then only $\rho_y$
asymptotically approaches to a constant.
$w_y$ approaches to $-1$,
but does not cross over $-1$.
The evolution is nearly similar to the non-coupling case.

Therefore,
for all four types of models that we have studied,
the coincidence problem can be naturally
solved by introducing the effective YMC as the dark energy
at the fixed parameter $\kappa$ given in Eq.(\ref{gamma}).
The present state of the universe is a natural result
of the dynamic evolution.
The past history of the evolving universe
is that of the standard Big Bang model,
and the future of the universe depends
on the details of the coupling.
If there is no coupling,
or if the matter decays,
or both matter and radiation decay,
into the YMC,  as in Model 2 and in Model 4,
the matter and the radiation will keep on decreasing
as  $\rho_m(t)\propto a(t)^{-3}$ and $\rho_r(t)\propto a(t)^{-4}$.
If the YMC decays into the matter only as in Model 1,
then $\rho_m(t)$ will asymptotically
remain as constant, like $\rho_y(t)$ does,
but $\rho_r(t)\propto a(t)^{-4}$.
If the YMC decays into both the matter and radiation
as in Model 3,
then all the components
$\rho_y(t)$,  $\rho_m(t)$, and $\rho_r(t)$
will asymptotically remain as constant.
In Model 1 and Model 3
the future of the universe is a steady state,
quite similar to that of the Steady State model,
thus, in a sense, these two models bridge
between the Big Bang model and the Steady State model.

The distinguished characteristics  of
the YMC dark energy  model are the following.

The YM field is known to be indispensable to particle physics,
the  effective YMC employed in our work
comes from quantum corrections up to 1-loop .
Therefore, there is no room to adjust
the form of the effective Lagrangian.
This is in contrast to scalar field models,
which have to design
the form of potential and sometimes even
the form of kinetic energy.

The solution of coincidence problem
has been relying on the parameter  $\kappa$ in all our models.
Viewed from the standard model of particle physics,
the energy scale by $\kappa$ is much smaller than the other known
microscopic energy scales.
And this stands as the fine-tuning problem
for any current cosmological model
so far, and for our model as well.
However, if the YM field in our model is regarded
as a fundamental gauge field
with $\kappa $ being the energy scale for this new physics,
then the fine-tuning problem is traced up to the new physics.
When the couplings are included,
there are two more parameters $\Gamma$ and $\Gamma'$.
But the present state of the universe requires
that $\Gamma$ be roughly the same order of magnitude of
the expansion rate $\Gamma\sim  0.5 H$,
and $\Gamma'$ be roughly three order lower.
Since $H\sim \sqrt{ G\kappa^2} = \kappa/m_{pl}$
with $m_{pl}$ being the Planck mass,
thus the couplings
$\Gamma\sim \kappa/m_{pl}$ and
$\Gamma' \sim 10^{-3} \kappa/m_{pl}$
are also associated with
the scale $\kappa $.

On the dynamic evolution, in comparison with scalar models,
our models have the following features.
All our models,
for a broad range of the initial condition
and for a variety of the coupling forms,
automatically have
the scaling property, i.e.,
$\rho_y(t)$ is initially subdominant to,
and keeps track of to the matter.
The accelerating stage begins only quite
recently around  a redshift $z\sim (0.35, \, 0.48)$.
This will allow the Big Bang cosmology to remain
without drastic modifications.
Besides, all our models  have only one stable fixed point,
uniquely determined by the ratio $\Gamma/H$
and has nothing to do with the initial conditions $\Omega_{yi}$.
Moreover, all the  quantities in our model,
especially $\rho_y(t)$ and $p_y(t)$
are continuous functions of $t$.
So there is no Big Rip singularities
in our models.
Interestingly,  as a function of $t$,  $w_y$
 behaves
quite smoothly during the evolution,
going from $\sim 1/3$, approaching to $-1$.
And in the case of the YMC decaying,
$w_y$ crosses over $-1$ at $z\sim 2$,
acquires the present value $\sim -1.1$,
and settles down to an asymptotic value $\sim -1.17$.

~~

ACKNOWLEDGMENT:
Y. Zhang's research work has been supported by
the Chinese NSF (10173008), NKBRSF (G19990754), and by SRFDP.
W. Zhao has been partially supported by Graduate Student Research
Funding from USTC.
We thank Dr. S. Tsujikawa for interesting  discussions.


\begin{figure}
\caption{\label{fig1} Non-coupling case: The evolution of energy
densities. For a wide range of initial conditions
$\rho_{yi}=(10^{-10}, \, 10^{-2})\rho_{mi}$, there always exists a
scaling solution during the early stages, and $\rho_y(t)$ levels
off and becomes dominant around $z\sim 0.35$.
 }
\end{figure}
\begin{figure}
\caption{\label{fig2} Non-coupling case: The evolution of fraction
of energy densities. The two curves of $\Omega_y$ for $y_i=1$ and
$y_i=16.12$ are very close to each other, and nearly overlap in
this figure.}
\end{figure}
\begin{figure}
\caption{\label{fig3}
Non-coupling case:
The evolution of EoS.
$w_y$ approaches to $-1$ as $t\rightarrow \infty$,
but does not cross over $-1$. }
\end{figure}

\begin{figure}
\caption{\label{fig4}
Model 1 with $Q_m>0$ for $\Gamma/H=0.5$ :
The evolution of energy densities
with the YMC decaying into the matter.
For a wide range of initial conditions
$\rho_{yi}=(10^{-10}, \, 10^{-2})\rho_{mi}$,
i.e., $y_i = (1, \, 15)$,
there always exists
a scaling solution during the early stages,
and $\rho_y(t)$ levels off and becomes
dominant around $z\sim 0.48$.
Due to the coupling, $\rho_m(t)$ also levels off at late time. }
\end{figure}
\begin{figure}
\caption{\label{fig5} Model 1 with $Q_m>0$ for $\Gamma/H=0.5$ :
The evolution of fractional energy densities in the same model as
in Fig.\protect\ref{fig4}. }
\end{figure}
\begin{figure}
\caption{\label{fig6} Model 1 with $Q_m>0$ for $\Gamma/H=0.5$ :
The evolution of EoS in the same model as in
Fig.\protect\ref{fig4} and  Fig.\protect\ref{fig5}. Due to the
coupling, $w_y$ crosses over $-1$, takes on value $\sim -1.1$ at
$z=0$, and approaches $-1.17$ asymptotically. This is in contrast
to the non-coupling case.}
\end{figure}

\begin{figure}
\caption{
\label{fig7} Model 1 with $Q_m>0$ for
$\Gamma/H=0.25e^{\frac{y}{4}}\frac{2+y}{1+y}$ : The evolution of
energy densities with the YMC decaying into the matter. For a wide
range of initial conditions $\rho_{yi}=(10^{-10}, \,
10^{-2})\rho_{mi}$, there always exists a scaling solution during
the early stages, and $\rho_y(t)$ levels off and becomes dominant
around $z\sim 0.48$. Due to the coupling, $\rho_m(t)$ also levels
off at late time. This is quite similar to the case of a constant
rate $\Gamma/H=0.5$ in Fig.\protect\ref{fig4}. }
\end{figure}
\begin{figure}
\caption{\label{fig8}.
Model 1 with $Q_m>0$ for $\Gamma/H=0.25e^{\frac{y}{4}}\frac{2+y}{1+y}$ :
The evolution of fractional energy densities in the same model
as in Fig.\protect\ref{fig7}.}
\end{figure}
\begin{figure}
\caption{\label{fig9}
Model 1 with $Q_m>0$ for $\Gamma/H=0.25e^{\frac{y}{4}}\frac{2+y}{1+y}$ :
The evolution of EoS  in the same model
as in Fig.\protect\ref{fig7} and Fig.\protect\ref{fig8}.
Again, due to the coupling, $w_y$ crosses over $-1$,
and takes on value $\sim -1.1$ at $z=0$.}
\end{figure}

\begin{figure}
\caption{\label{fig10}
Model 2 with  $Q_m<0$ for the constant coupling $\Gamma/H=0.02$ :
The evolution of energy densities with
the matter decaying into the YMC.
 }
\end{figure}
\begin{figure}
\caption{\label{fig11}
Model 2 with  $Q_m<0$ for the constant coupling $\Gamma/H=0.02$ :
The evolution of fractional energy densities in the same model
as in Fig.\protect\ref{fig10}.
}
\end{figure}
\begin{figure}
\caption{\label{fig12}
Model 2 with  $Q_m<0$ for the constant coupling $\Gamma/H=0.02$ :
The evolution of EoS  in the same model
as in Fig.\protect\ref{fig10} and Fig.\protect\ref{fig11}.
Since the YMC gets energy from the matter,
$w_y$ approaches to $-1$,  but does not cross over $-1$.}
\end{figure}

\begin{figure}
\caption{\label{fig13}
 Model 3  with $Q_m>0$ and $Q_r>0$
 for $\Gamma/H=0.5$ and $\Gamma'/H=0.00018$ :
The evolution of energy densities with
the YMC decaying into both
the matter and the radiation.
For a wide range of initial conditions
$\rho_{yi}=(10^{-10}, \, 10^{-2})\rho_{mi}$,
there always exists a scaling solution during the early stages,
and $\rho_y(t)$ levels off and becomes
dominant around $z\sim 0.48$.
Note that,  due to coupling,
both $\rho_m(t)$  and $\rho_r(t)$ level off
like $\rho_y(t)$.}
\end{figure}

\begin{figure}
\caption{\label{fig14}
The evolution of fraction of energy densities in
the same model as in  Fig.\protect\ref{fig13}. }
\end{figure}
\begin{figure}
\caption{\label{fig15}
 Model 3  with $Q_m>0$ and $Q_r>0$
 for $\Gamma/H=0.5$ and $\Gamma'/H=0.00018$ :
The evolution of EoS in
the same model as in  Fig.\ref{fig13} and Fig.\protect\ref{fig14}.
Note that $w_y$ also crosses over $-1$
and takes on a value $\sim -1.1$ at $z=0$. }
\end{figure}

\begin{figure}
\caption{\label{fig16}
Model 1 with $Q_m>0$ for $\Gamma/H=0.5$ :
The trajectories in the phase plane.
Each trajectory starts with a different initial condition.
All of them approach the fixed point
$(x_f, y_f)=( 0.13666, \,  -0.11499 )$. }
\end{figure}

\begin{figure}
\caption{\label{fig17}
Model 3  with $Q_m>0$ and $Q_r>0$
 for $\Gamma/H=0.5$ and $\Gamma'/H=0.00018$ :
The trajectories in the phase plane $(x,y)$.
Each trajectory starts
with a different initial condition.
All of them approach the fixed point
$(x_f, ~ y_f, ~ r_f) = (0.13666,~ -0.11503, ~ 4\times10^{-5})$.
The parameters are the same as in Fig.\protect\ref{fig13}. }
\end{figure}

\begin{figure}
\caption{\label{fig18}
Model 3  with $Q_m>0$ and $Q_r>0$
 for $\Gamma/H=0.5$ and $\Gamma'/H=0.00018$ :
The trajectories in the phase plane $(r,y)$.
The parameters are the same as in Fig.\protect\ref{fig13}.   }
\end{figure}

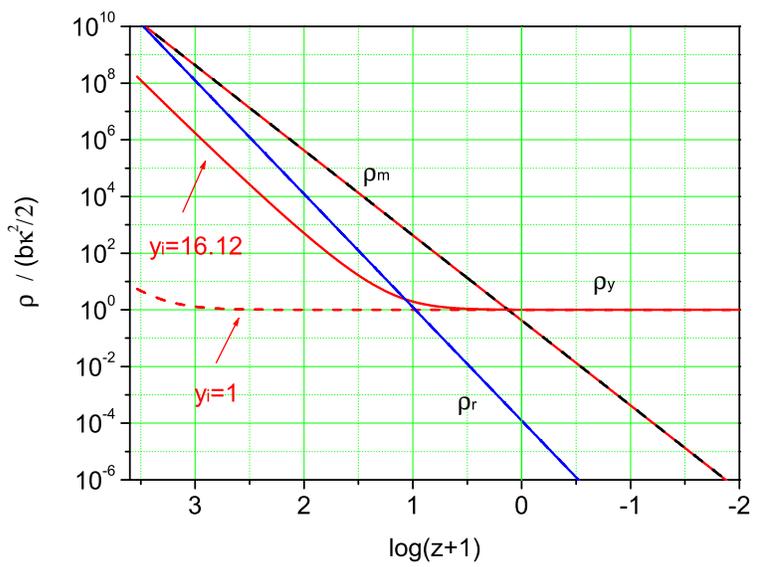
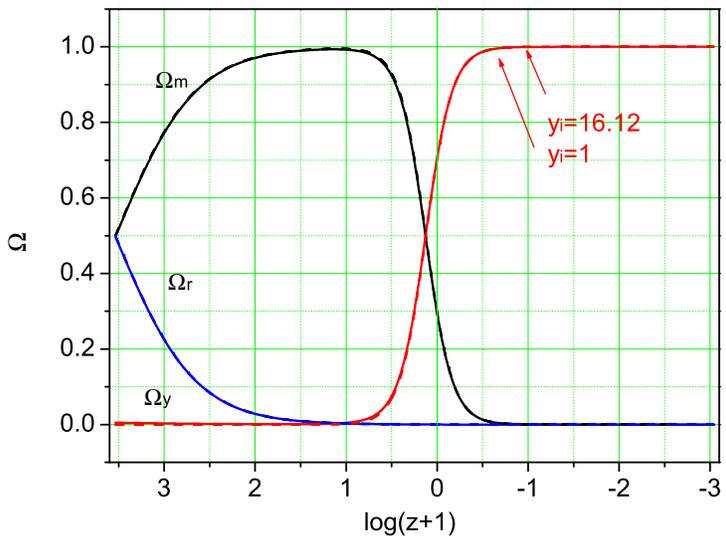
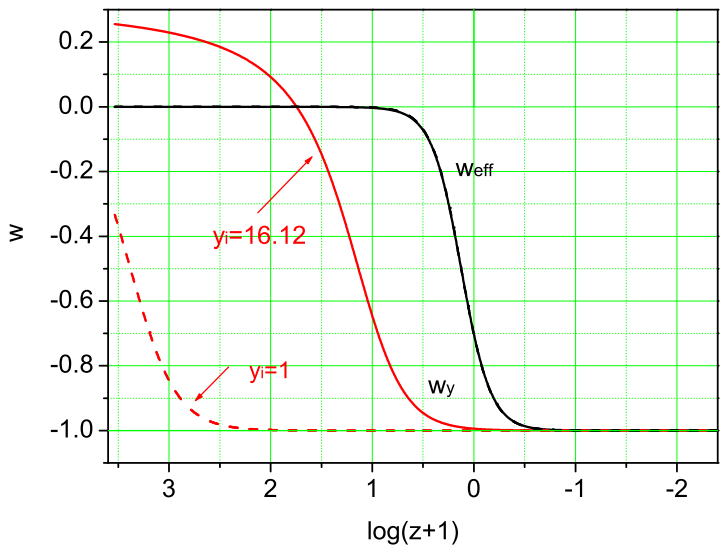
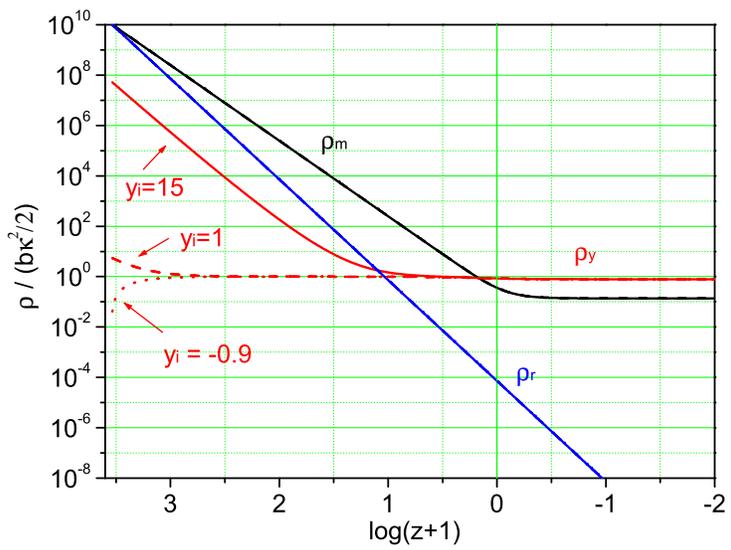
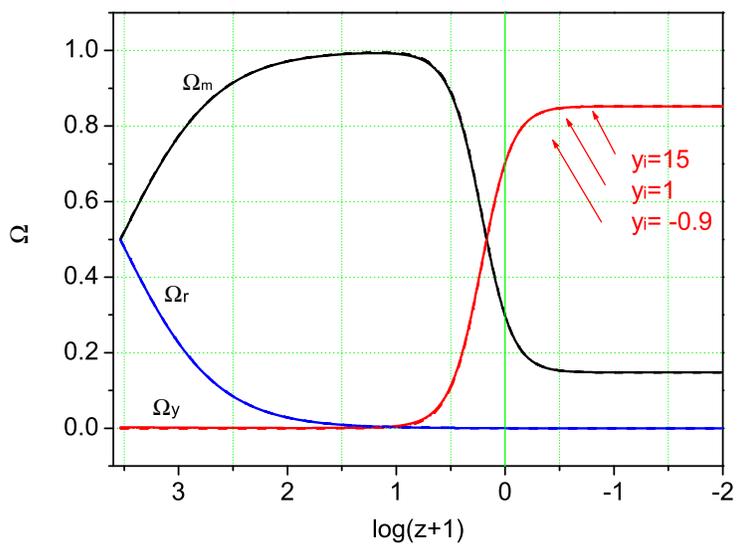
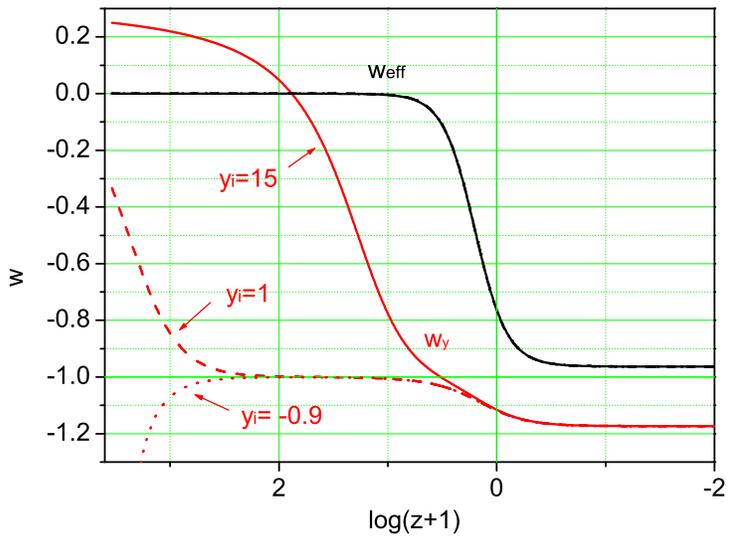
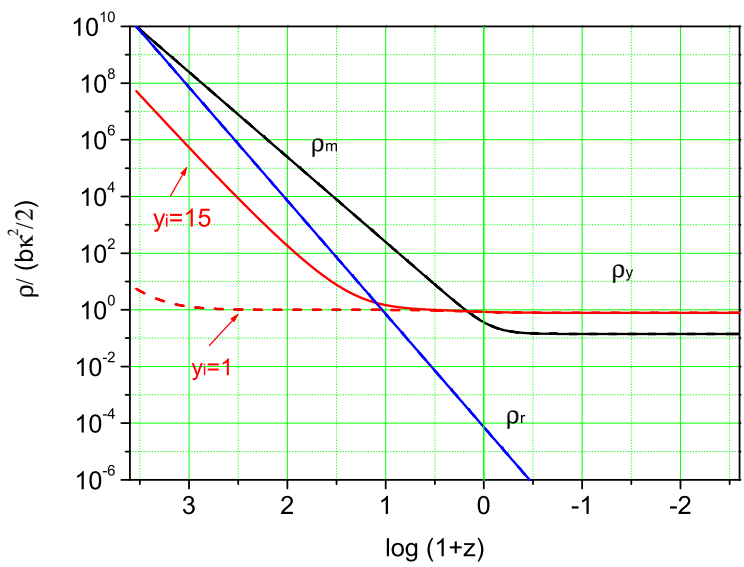
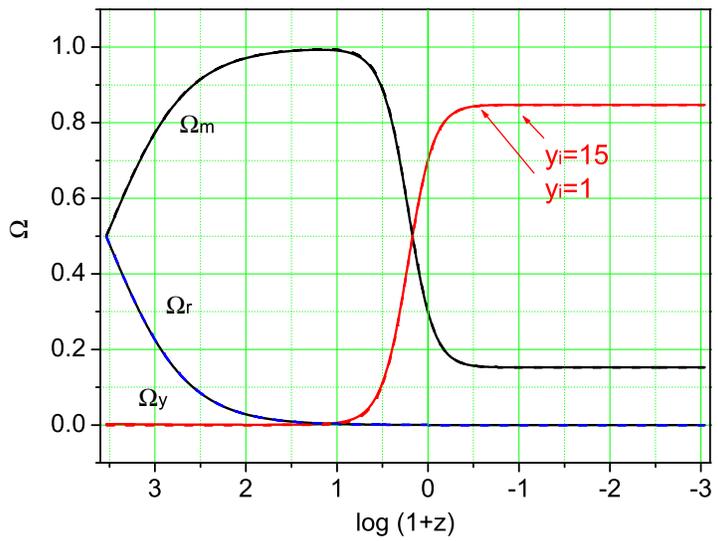
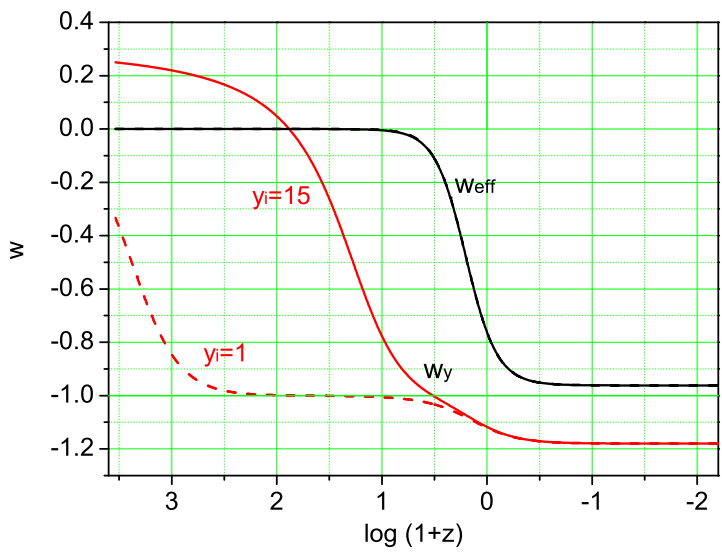
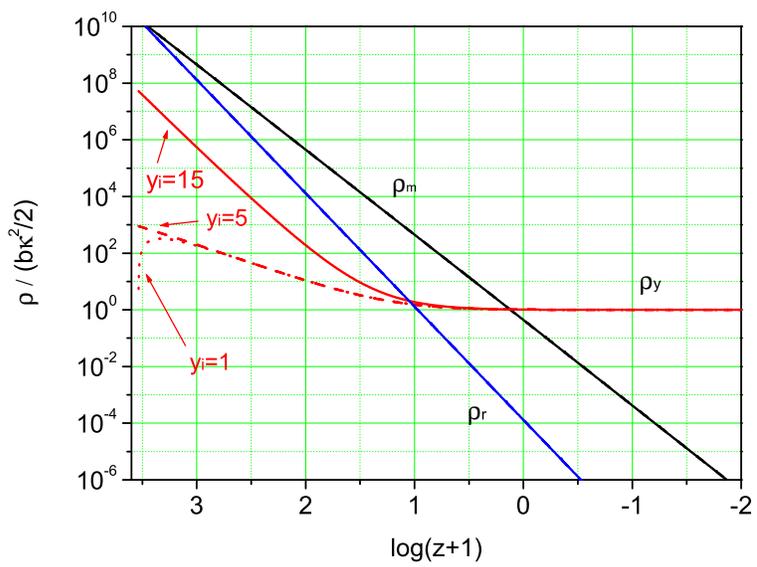
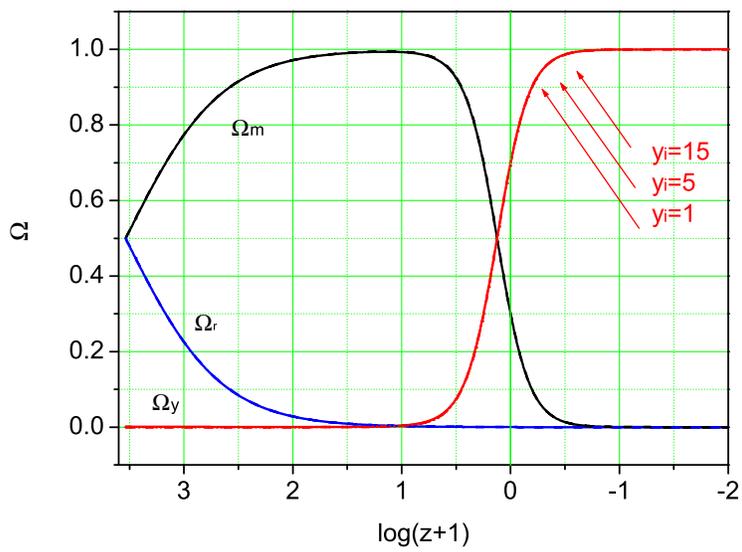
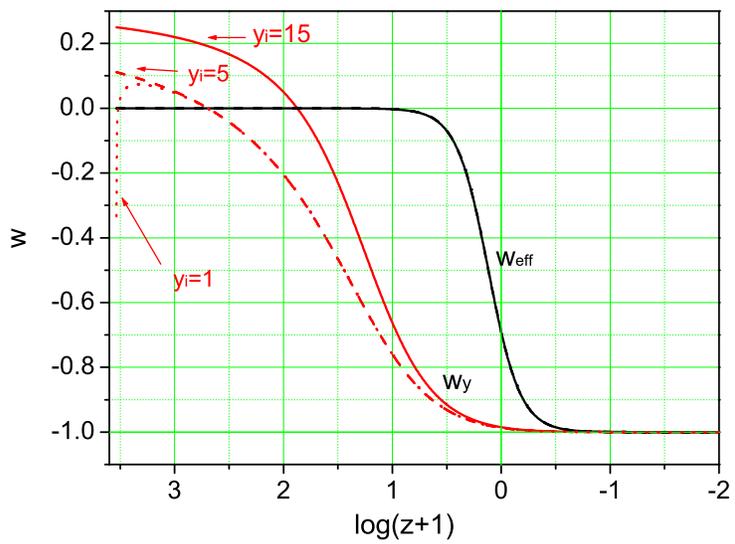
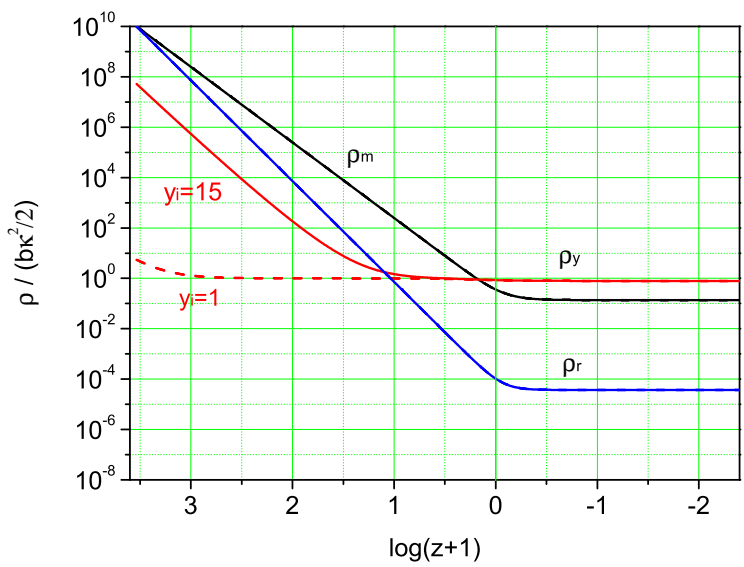
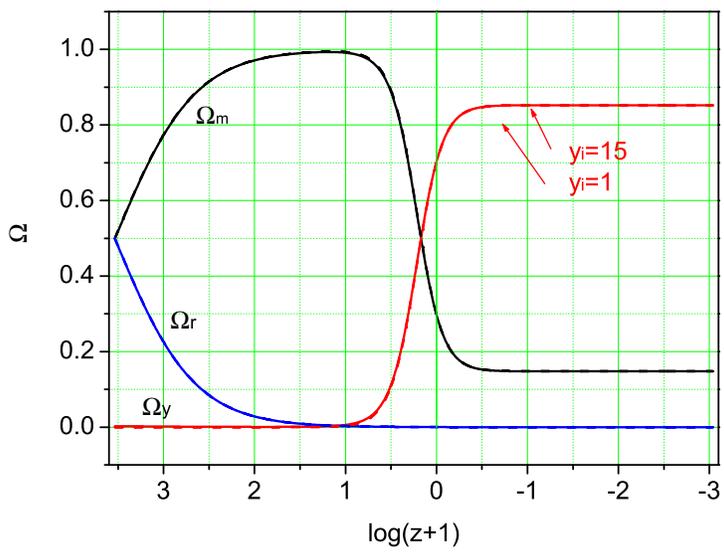
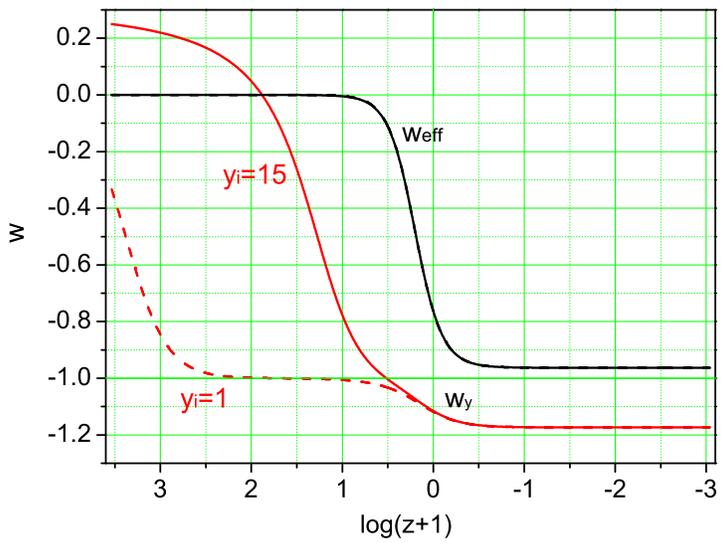
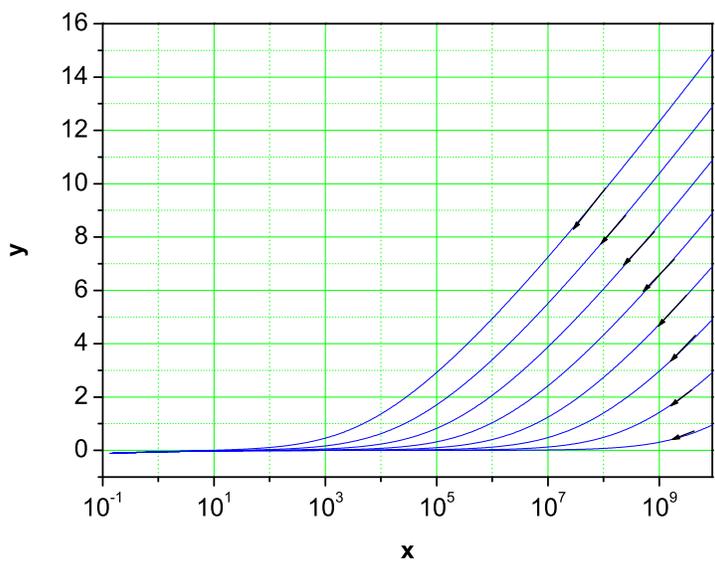
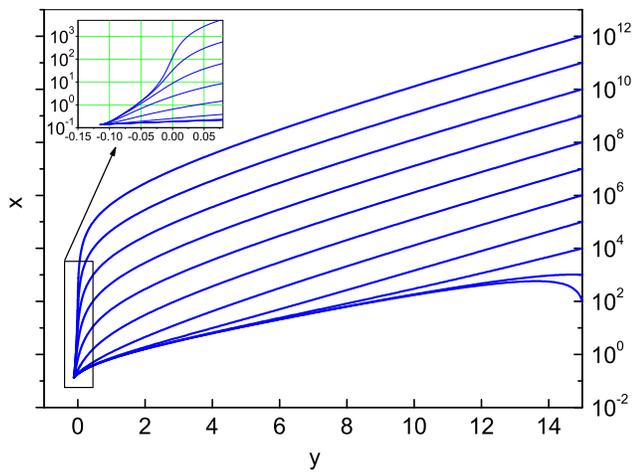
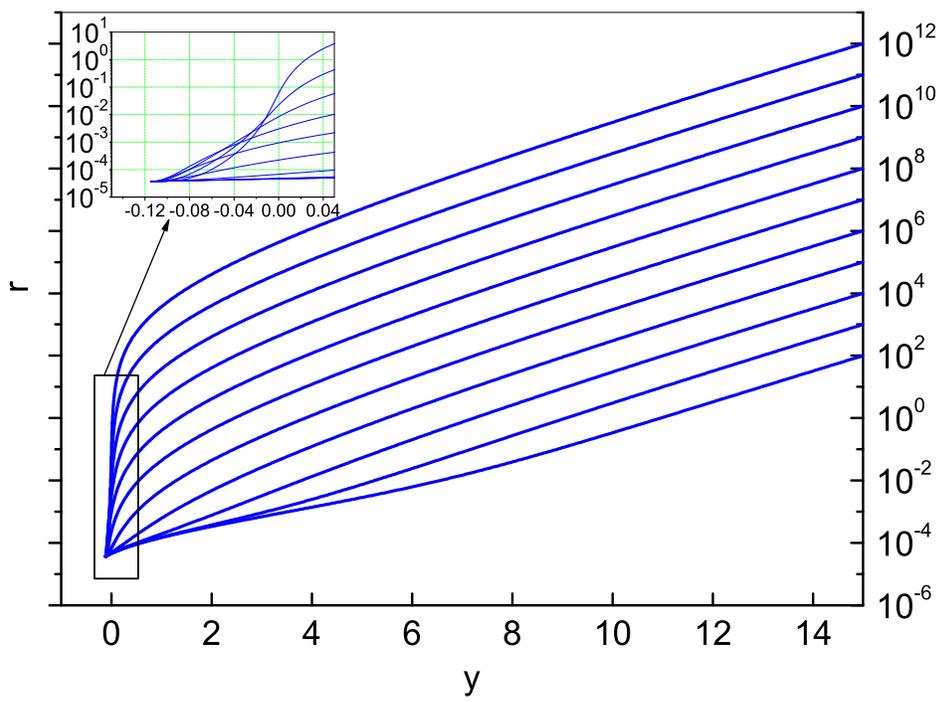
\end{document}